\documentclass[12pt]{article}

\RequirePackage{amsmath,amsthm,amsfonts,amssymb}
\usepackage{hyperref}
\RequirePackage{graphicx}
\usepackage[dvipsnames]{xcolor}
\usepackage{float}
\usepackage{enumitem}
\usepackage{url}
\usepackage{bbm}
\usepackage{natbib}
\usepackage[nottoc,numbib]{tocbibind}
\usepackage{booktabs,caption,subcaption}
\usepackage[flushleft]{threeparttable}
\usepackage{multirow}

\usepackage{xr}
\usepackage{color}
\usepackage{tikz}
\usetikzlibrary{patterns}
\usepackage{mathtools}
\usepackage{booktabs}
\usepackage{multirow}
\usepackage{makecell}
\usepackage{colortbl}
\usepackage{siunitx}
 \usepackage{gensymb}
     
\usepackage[margin=1in]{geometry}

\newcommand{\bA}{\boldsymbol{A}}

\newcommand{\bb}{\boldsymbol{b}}
\newcommand{\bB}{\boldsymbol{B}}

\newcommand{\mD}{\mathcal{D}}

\newcommand{\bh}{\boldsymbol{h}}

\newcommand{\bj}{\boldsymbol{j}}
\newcommand{\bJ}{\boldsymbol{J}}

\newcommand{\bq}{\boldsymbol{q}}
\newcommand{\bs}{\boldsymbol{s}}

\newcommand{\mS}{\mathcal{S}}

\newcommand{\bv}{\boldsymbol{v}}

\newcommand{\bW}{\boldsymbol{W}}

\newcommand{\bzero}{\boldsymbol{0}}

\newcommand{\bPsi}{\boldsymbol{\Psi}}
\newcommand{\bpsi}{\boldsymbol{\psi}}

\newcommand{\btheta}{\boldsymbol{\theta}}

\theoremstyle{definition}

\usepackage[draft,inline,nomargin,index]{fixme}
\fxsetup{theme=color,mode=multiuser}
\FXRegisterAuthor{al}{aal}{\color{purple} AL}
\FXRegisterAuthor{eh}{aeh}{\color{PineGreen}EH}
\FXRegisterAuthor{rh}{arh}{\color{RoyalBlue}RH}

\allowdisplaybreaks

\begin{document}

\def\spacingset#1{\renewcommand{\baselinestretch}%
{#1}\small\normalsize} \spacingset{1}

\title{\bf When the whole is greater than the sum of its parts: Scaling black-box inference to large data settings through divide-and-conquer}
\author{Emily C. Hector\\Department of Statistics, North Carolina State University\\
Amanda Lenzi\\School of Mathematics, University of Edinburgh}
\date{}
\maketitle
 
\bigskip
\begin{abstract} 
Black-box methods such as deep neural networks are exceptionally fast at obtaining point estimates of model parameters due to their amortisation of the loss function computation, but are currently restricted to settings for which simulating training data is inexpensive. When simulating data is computationally expensive, both the training and uncertainty quantification, which typically relies on a parametric bootstrap, become intractable. We propose a black-box divide-and-conquer estimation and inference framework when data simulation is computationally expensive that trains a black-box estimation method on a partition of the multivariate data domain, estimates and bootstraps on the partitioned data, and combines estimates and inferences across data partitions. Through the divide step, only small training data need be simulated, substantially accelerating the training. Further, the estimation and bootstrapping can be conducted in parallel across multiple computing nodes to further speed up the procedure. Finally, the conquer step accounts for any dependence between data partitions through a statistically and computationally efficient weighted average. We illustrate the implementation of our framework in high-dimensional spatial settings with Gaussian and max-stable processes. Applications to modeling extremal temperature data from both a climate model and observations from the National Oceanic and Atmospheric Administration highlight the feasibility of estimation and inference of max-stable process parameters with tens of thousands of locations.
\end{abstract}

\noindent%
{\it Keywords:} 
Amortised inference, Convolutional neural networks, Gaussian process, Generalized method of moments, Max-stable process, Statistical computing.

\section{Introduction}\label{s:intro}

\subsection{Background on black-box inference}

Modeling spatial datasets is computationally challenging, even with restrictive model assumptions and in moderate dimensions.
Consider Gaussian processes, a prevalent model choice in spatial statistics due to its mathematical and computational convenience such as Gaussian marginal and conditional distributions. Minimizing the negative log-likelihood of these processes directly is prohibitively expensive even for a moderate number of spatial locations $d$ because determinant operations have cubic time complexity. Various methods have been proposed over the years to overcome this computational challenge. \citet{Heaton-etal-2019} grouped these methods into lowrank, sparse covariance matrices, sparse precision matrices and algorithmic. Lowrank and sparse covariance approaches simplify the $d \times d$ covariance matrix by either reducing its rank \citep{cressie2006spatial, cressie2008fixed, solin2020hilbert}, by encoding sparsity through the introduction of zeros in some of its entries through tapering \citep{furrer2016asymptotic} or by selecting only a suitable number of neighbours in the matrix calculations \citep{vecchia1988estimation, stein2004approximating, datta2016hierarchical, katzfuss2020class}. An alternative approach is based on Gaussian Markov random field approximations, which successfully take advantage of the natural sparsity in precision matrices instead of using the covariance matrix \citep{lindgren2011explicit}. Algorithmic approaches are a less well defined category focused on fitting schemes rather than model building \citep[see e.g.][]{gramacy2015local}. 

Traditional Gaussian processes are not only computationally infeasible for large spatial data but also too simplistic to describe complex dependencies. The large amount of data over extensive spatiotemporal domains that arise from recent advances in data collection and storage require more complex models in order to accurately characterize their dependence structures. The use of more complicated models, however, often comes with the cost of log-likelihoods that are intractable without a closed form. In response to these computational challenges, various simulation-based methods have been proposed, such as Approximate Bayesian computation (ABC) \citep{sisson2018overview}. These methods, often also refered to as likelihood-free, aim to perform inference for intractable likelihoods using simulations from the model, which are computationally tractable and inexpensive, therefore bypassing the log-likelihood evaluation altogether. The main idea is to identify the model parameters that yield simulated data that resemble the observed data \citep{cranmer2020frontier}.

More recently, in response to the widespread success of and easy access to deep learning techniques, new likelihood-free statistical inference methods have emerged that learn a mapping between the sample space and the parameter space using deep neural networks \citep{gerber2021fast, lenzi2021neural, sainsbury2022fast, sainsbury2024neural, jordan2023neural, lenzi2023towards, maceda2024variational, zammit2024neural, feng2024amortized, sainsbury2024likelihood}. These methods have been successful with Gaussian and max-stable process models for spatial data. The idea is to train a neural network on parameter values and corresponding data simulated from these parameter values. After an initial investment in generating simulated data and training the neural network, inference for new data can be performed without further simulations, as long as they are of the same size as the training data. After training, the neural network is fit to the observed data to obtain point parameter estimates, and inference relies on bootstrapping from the fitted model to estimate standard errors and confidence intervals. The inference is said to be \emph{amortised} because the fitted neural network can be used to obtain estimates from any new observed data of the same size at a much reduced computational cost compared to classical full or approximate likelihood methods.

While training neural networks to fit to new observed data is relatively fast, some models, such as the max-stable process, remain computationally challenging with large spatial domains due to the expense of generating large amounts of training data and bootstrapping. For this reason, existing black-box estimation methods with max-stable processes consider at most $d = 700$ locations. Moreover, once trained on a spatial domain of $d$ locations, for most methods, the neural network cannot be used to obtain estimates and inference for observed data with a different number of spatial locations. The simulation of training data and the training itself must be begun anew to carry out inference on observed data of a different size.

Therefore, despite the development of these amortised methods, parameter estimation in large spatial settings remains computationally expensive. The primary goal of this paper is to present a computationally and statistically efficient framework for black-box parameter estimation and inference with large spatial domains. 

\subsection{Our contributions}

Our proposal to achieve this goal is to leverage the divide-and-conquer framework, whereby a large spatial domain is divided into blocks of equal size, the model is fit in each block separately and in parallel, and the point estimates and measures of their uncertainty are combined using a computationally and statistically efficient rule. The divide-and-conquer approach we develop, based on black-box parameter inference, drastically reduces the computational complexity of black-box inference methods with large spatial domains. It does so in two ways: (i) data are quickly simulated on a small spatial domain, also accelerating the neural network training; and (ii) estimation and inference are amortised in the sense that a neural network trained on data of size $d_1$ can be used for inference on data of size $d_2$, where $d_1$ is not necessarily equal to $d_2$.

This framework is especially computationally appealing because training data are only as large as the small blocks and so we need only simulate data on a small domain. In contrast, other simulation-based methods usually require simulation from the selected spatial process on the entire domain, which is not computationally feasible in high-dimensional settings. This further means that our framework is scalable to very high dimensions even when fast simulation from the model is not possible, broadening the class of models which can be fit using black-box methods. Besides computational efficiency during training and fitting, another major benefit of training the neural network on small blocks is that, once the neural network has been trained, estimation is independent of the actual data size. While most existing deep learning methods are amortised in the sense that the trained neural network can be re-used for multiple data sets at almost no computational cost, the amortisation is only valid provided that each data set has the same format as those used to train the neural network. Our inference framework is fully amortised because the trained neural network can be used on much larger domains than the training domain.

The key challenge in developing our divide-and-conquer approach is in combining the dependent parameter estimates from each block into a global estimate with calibrated uncertainty quantification such that, for example, confidence intervals reach their nominal coverage. Most (pseudo)likelihood-based divide-and-conquer methods focus on prediction due to this difficulty \citep[see, e.g.][]{Lee-Park}. Recently, \cite{Hector-Reich, Hector-Reich-Eloyan} extended the combination rule of \cite{Hector-Song-JASA, Hector-Song-JMLR} to the spatial setting. Their approaches, respectively designed for max-stable and Gaussian processes, remain computationally burdensome when the number of spatial locations $d$ is large because they rely on composite likelihood and full likelihood estimation, respectively, within each block. Further, they both require repeated, independent observations of the spatial domain to estimate the dependence between estimators from each block, which may be unavailable in some spatial settings. Our proposal, described in Section \ref{s:dac}, is to replace the (peusdo)likelihood evaluation in each block with the black-box parameter estimation, and to use the bootstrap estimates from the small blocks to estimate the dependence between blocks. 
Although confidence intervals are essential for rigorous inference assessment and for reporting results that others can trust and replicate, previous work in black-box model fitting has mostly focused on point estimation. In part, this is because confidence intervals based on black-box parameter inference have poor coverage rates, as discussed in Section S2.7 of the supplement in \cite{zammit2024neural}. This phenomenon is due to the discrepancy between the trained estimator and the optimal estimator (that minimizes the expected loss), referred to as the ``amortisation gap'' \citep{cremer2018inference, zammit2024neural}. In this paper, we propose selecting a neural network from multiple trained networks to minimize the influence of the amortisation gap on the downstream inference. We illustrate this empirical strategy both without and with the divide-and-conquer framework described above in Sections \ref{s:uncertainty-bb} and \ref{sex:simu}, respectively.

Finally, in Section \ref{s:data} we provide and discuss two different applications, from climate model reanalysis and historical data, that fit Brown-Resnick max stable processes to $d=32{,}400$ spatially dependent annual temperature maxima. These processes are well-known for having likelihoods that are intractable even at moderate dimensions, and to our knowledge existing work on neural estimation is limited to a maximum of $d=900$ locations, although this is a rapidly evolving field. The results of these applications demonstrate that there can be substantial gains to our divide-and-conquer procedure. 
Our method provides reasonable estimates locally and globally, is statistically and computationally efficient, and easy to implement, providing strong support its use in practice.

\section{Black-box inference}\label{s:uncertainty-bb}

\subsection{Black-box estimation}

Let $\mS \subset \mathbb{R}^2$ a spatial domain and $Y(\bs)$ the outcome value at location $\bs \in \mS$. Denote by $Y(\mS) = \{Y(\bs): \bs \in \mS\}$ the random field. Suppose we observe $Y(\mS)$ at a set of gridded locations $\mD=\{\bs_j\}_{j=1}^d$, denote by $Y(\mD) = \{Y(\bs_j): \bs_j \in \mD\}$, and let $f\{ y(\mD); \btheta_0\}$ be the joint density of $Y(\mD)$ parameterized by a vector $\btheta \in \mathbb{R}^q$ with true value $\btheta_0$. We first investigate the amortisation gap of neural network training when used to fit models with no divide-and-conquer strategy, and so we assume for now that the number of locations $d$ is not too large. This investigation will point towards strategies for accurately estimating the uncertainty of point estimates that will be used in the divide-and-conquer strategy described in Section \ref{s:dac}. In our examples below, we consider two spatial processes to illustrate the implementation of black-box model fitting: a Gaussian and a max-stable process. 

The main idea to estimate $\btheta$ based on the observed data $Y(\mD)$ is to simulate data from the model to train a neural network that can be used to fit the model to $Y(\mD)$. If $\{\btheta_t\}_{t=1}^T$ is a set of candidate parameter values, we generate training outcomes $Y_t(\mD)$ from $f\{ y(\mD); \btheta_t\}$, $t=1, \ldots, T$ to generate training data pairs $\{\btheta_t, Y_t(\mD)\}_{t=1}^T$. Ideally, $\btheta_t$ are close to the true value $\btheta_0$. On the one hand, these values will be used to generate training values of our outcome, so they should be relatively ``close'' to the true values. This can be achieved by generating training values around an initial guess of the value of $\btheta_0$, for example based on a computationally inexpensive initial estimate of $\btheta$ in $\mD$. On the other hand, they should have a large enough variance to generate simulated outcomes that are sufficiently variable. This is particularly important to ensure that our black-box algorithm can learn to discriminate data generated by different parameter values. These values can be randomly distributed according to a chosen distribution or deterministically generated on a grid. We take the latter approach in our numerical illustrations by choosing $\btheta_t$ on a grid.

A convolutional neural network (CNN) is then trained using the simulated data $\{Y_t(\mD)\}_{t=1}^T$ as inputs and the parameter values $\{\btheta_t\}_{t=1}^T$ as outputs to learn the map between observed data and the generating parameter values; in the case of the mean squared error (MSE) loss, the estimated map minimizes the conditional expectation of parameters given data. An example architecture of a deep CNN with three two-dimensional convolutional layers is given below in Table \ref{tab:cnn-arch}. Denote by $a(\widehat{\bA}; \cdot)$ the network with input $\cdot$ and trained weights $\widehat{\bA}$ that minimize the training squared error loss: $\widehat{\bA} = \arg \min_{\bA} \sum_{t=1}^T [\btheta_t - a\{ \bA; Y_t(\mD)\}]^2$. The parameter estimate is obtained using $\widehat{\btheta}=a\{ \widehat{\bA}; Y(\mD) \}$. The sampling distribution of $\widehat{\btheta}$ is estimated using $B$ bootstrap replicates $\widehat{\btheta}_b=a \{ \widehat{\bA} ; Y_b(\mD)\}$ of $\widehat{\btheta}$, where $Y_b(\mD)$ is sampled from $f\{ y(\mD); \widehat{\btheta}\}$, $b=1, \ldots, B$.

\begin{table}[ht!]
\centering
\begin{tabular}{ccccc} 
Layer Type & Filters/units & Kernel size &  Max pool size & Activation \\ 
\midrule
2D conv &  128 & $10 \times 10$ & $2 \times 2$ & ReLU \\ 
2D conv & 128 & $5 \times 5$ &  $2 \times 2$ & ReLU\\ 
2D conv &  128 & $3 \times 3$ & $2 \times 2$ & ReLU \\ 
dense & 500 &  &  & ReLU \\ 
dense & 2 &  & & linear \\ 
\midrule
Total trainable weights: & & & $827{,}742$
\end{tabular}
\caption{Summary of the CNN model architecture. For all two-dimensional convolutional layers, we set \texttt{leak} = 0.1 in the ReLU activation function and \texttt{padding = `same'} for the max pooling layers. 
The optimization parameters are: \texttt{learning rate} = 0.001, \texttt{loss} = `mse', \texttt{number of epochs} = 300,  \texttt{batch size} = 200 with early stopping, \texttt{patience} = 50. \label{tab:cnn-arch} }
\end{table}

\subsection{Uncertainty quantification}\label{ss:UQ}

We illustrate with a simulation the influence on inference of the stochastic nature of gradient descent variants, non-deterministic operations due to parallel processing and differences in floating-point calculations that accumulate over many iterations. The outcome $Y(\mD)$ is generated from a mean-zero Gaussian distribution with covariance function
\begin{align}
C\{ y(\bs_j), y(\bs_{j^\prime}) \} &= \tau^2_0 \exp(-\|\bs_j - \bs_{j^\prime} \|_2/\phi^2_0),
\label{e:gaussian-covariance}
\end{align}
with $\tau^2_0>0$ and $\phi^2_0>0$ the true variance and correlation parameters, respectively, and $\bs_j,\bs_{j^\prime} \in \mD$. We take $\mD=[1,20]^2$ to be a square gridded spatial domain of dimension $d=20\times 20$. The true values of the spatial covariance are $\tau^2_0=3$ and $\phi^2_0=3$. Our aim is to estimate and make inference on $\btheta_0=\{ \log(\tau^2_0), \log(\phi^2_0)\}$.

Values of $\log(\phi^2_{t_1})$, $t_1=1, \ldots, T_1$, consist of the sequence from $\log(\phi^2_0)-0.5$ to $\log(\phi^2_0)+0.5$ (approximately $1.82$ to $4.95$) of length $T_1=150$, and values of $\log(\tau^2_{t_2})$, $t_2=1, \ldots, T_2$, consist of the sequence from $\log(\tau^2_0)-0.5$ to $\log(\tau^2_0)+0.5$ (approximately $1.82$ to $4.95$) of length $T_2=150$. Values $\btheta_t = \{ \log(\tau^2_{t_1}), \log(\phi^2_{t_2}) \}_{t_1,t_2=1}^{T_1,T_2}$, $t=1, \ldots, 22{,}500$, consist of all combinations of $\log(\tau^2_{t_1})$ and $\log(\phi^2_{t_2})$. For each $\btheta_t$, we simulate $Y_t(\mD)$ on $\mD$ from the model. The training and observed values of the outcome are transformed using $\mathbbm{1}(y >0)\log(y) -\mathbbm{1}(y \leq 0) \log(-y)$. The training values of $\btheta$ are centered by the minimum training value and standardized by the range of the training values. The neural network architecture is described in Table \ref{tab:cnn-arch}. A validation sample is generated similarly to the training sample but with $T_1=T_2=70$. We train the network 500 times using the same architecture and training and validation data but 500 different seeds. For comparison, we also estimate $\btheta=\{ \log(\tau^2), \log(\phi^2)\}$ using maximum likelihood estimation (MLE), supplying the true value $\btheta_0$ as the starting value in the maximization of the log-likelihood.

We focus our discussion on five neural networks (all with the same architecture and training and validation data) corresponding to the neural networks with smallest and largest (in absolute value) average validation bias (AVB), and AVB at the tertiles (in absolute value). For each of these five trained neural networks $a_r\{\widehat{\bA}_r; Y(\mD)\}$, $r=1, \ldots, 5$, sorted from smallest to largest AVB, we obtain a point estimate of $\btheta_0$ using $\widehat{\btheta}_r = a \{ \widehat{\bA}_r; Y(\mD)\}$. We estimate the sampling distribution of $\widehat{\btheta}_r$ using $B=5{,}000$ bootstrap replicates $\widehat{\btheta}_{rb}=a\{\widehat{\bA}_r; Y_{rb}(\mD)\}$, where $Y_{rb}(\mD)$ is sampled from $f\{y(\mD); \widehat{\btheta}_r\}$, $r=1, \ldots, 5$, $b=1, \ldots, B$. 

\begin{figure}[h!]
\centering
\includegraphics[width=\textwidth]{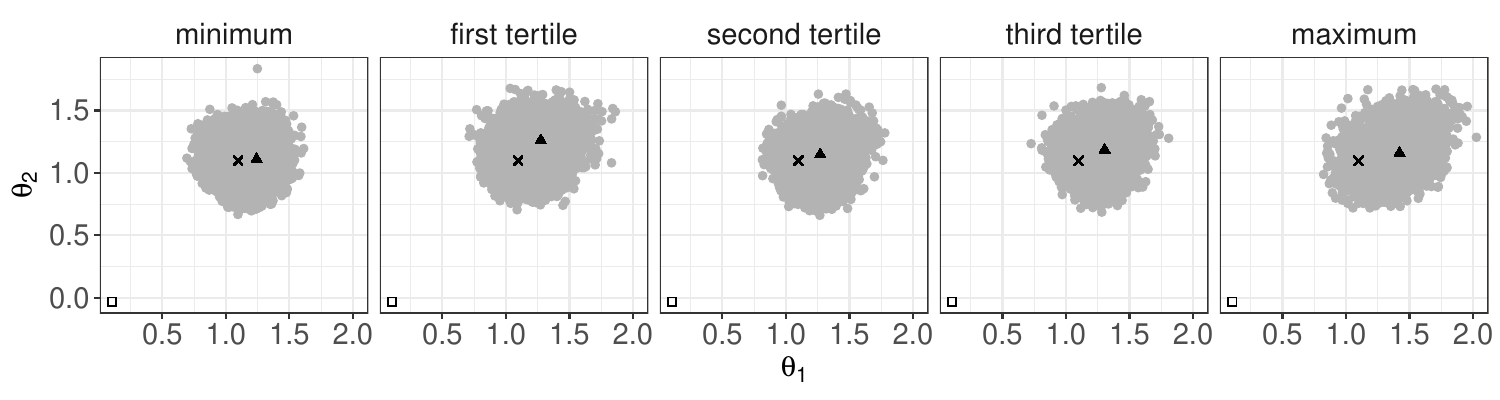}
\caption{Neural network estimates (triangles), maximum likelihood estimates (squares), true values (crosses) and bootstrap replicates (grey dots) of $\btheta_0$ based on the five trained neural networks with minimum, first, second, third tertile and maximum AVB. \label{f:small_sim_bootstrap}}
\end{figure}

Figure \ref{f:small_sim_bootstrap} plots the point estimates $\widehat{\btheta}_r$, the bootstrap replicates $\widehat{\btheta}_{rb}$, the maximum likelihood estimator and the true value $\btheta_0$ for each of the $r=1, \ldots, 5$ neural networks. While the point estimates are close to the true value $\btheta_0$ across the five neural networks, the bias appears to increase somewhat as the AVB increases. The variance of the bootstrap samples also appears to increase as the AVB increases. This seems to suggest that the neural network with larger AVB has not learned the map from data to parameters as well as the neural network with smaller AVB, a manifestation of the amortisation gap.

To validate these observations, we evaluate the frequentist properties of our point estimate $\widehat{\btheta}_r$ and its estimated sampling distribution based on the bootstrap replicates. We sample 500 observed outcomes $Y(\mD)$ as above, and for each observed outcome, we obtain $\widehat{\btheta}_r$ and an estimate of its sampling distribution using $B=5{,}000$ bootstrap replicates, $r=1, \ldots, 5$. From the estimated sampling distribution of $\widehat{\btheta}_r$, we estimate the standard error of $\widehat{\btheta}_r$ using the standard deviation of the bootstrap replicates. Table \ref{tab:small_sim_bootstrap} reports the average bias (BIAS), root mean squared error (RMSE) and the estimated standard error (SE) each averaged across the 500 Monte Carlo simulation replicates, as well as the empirical standard deviation (SD) of $\widehat{\btheta}_r$ across the 500 Monte Carlo simulation replicates. Across the 500 Monte Carlo replicates, the BIAS does indeed increase as the AVB increases. Further, the SE tracks the SD more closely for neural networks with smaller AVB, suggesting that $a\{\widehat{\bA}_1; Y(\mD)\}$ may be better capturing the inherent variability in the data. On the other hand, the MLE exhibits substantial bias.

The run time of estimating $\btheta_0$ using $a\{ \widehat{\bA}_1; Y(\mD)\}, \ldots, a\{ \widehat{\bA}_5; Y(\mD)\}$ is, respectively, $0.658, 0.491, 0.537, 0.494, 0.548$ seconds averaged across the 500 simulations. In comparison, computing the maximum likelihood estimator is approximately $17$ times slower, taking $9.24$ seconds averaged across the 500 simulations. This makes it computationally prohibitive to estimate the sampling distribution of the maximum likelihood estimator using a parametric bootstrap, and we therefore are unable to compute its standard deviation or confidence intervals.

\begin{table}[htb!]
\centering
\begin{tabular}{ccrrrr} 
method & parameter & BIAS$\times 100$ & RMSE$\times 10$ & SE$\times 10$ & SD$\times 10$ \\
\midrule
\multirow{2}{*}{$a\{\widehat{\bA}_1; Y(\mD)\}$} & $\phi^2$ & $-0.764$ & 1.37 & 1.40 & 1.37 \\ 
& $\tau^2$ & 0.674 & 1.43 & 1.46 & 1.43 \\ 
\multirow{2}{*}{$a\{\widehat{\bA}_2; Y(\mD)\}$} & $\phi^2$ & 6.03 & 1.67 & 1.62 & 1.55 \\ 
& $\tau^2$ & 1.68 & 1.44 & 1.48 & 1.44 \\ 
\multirow{2}{*}{$a\{\widehat{\bA}_3; Y(\mD)\}$} & $\phi^2$ & 10.8 & 1.79 & 1.50 & 1.43 \\ 
& $\tau^2$ & $-0.396$ & 1.44 & 1.51 & 1.44 \\ 
\multirow{2}{*}{$a\{\widehat{\bA}_4; Y(\mD)\}$} & $\phi^2$ & 10.5 & 1.75 & 1.45 & 1.40 \\ 
& $\tau^2$ & 2.39 & 1.46 & 1.51 & 1.45 \\ 
\multirow{2}{*}{$a\{\widehat{\bA}_5; Y(\mD)\}$} & $\phi^2$ & 14.3 & 2.15 & 1.74 & 1.62 \\ 
& $\tau^2$ & 6.61 & 1.72 & 1.63 & 1.58 \\ 
\multirow{2}{*}{MLE} & $\phi^2$ & $-12.7$ & 12.8 & 1.63 & -- \\ 
& $\tau^2$ & $-9.65$ & 9.73 & 1.31 & -- \\ 
\end{tabular}
\caption{Simulation metrics for the non-distributed neural estimator in the Gaussian simulations. \label{tab:small_sim_bootstrap}}
\end{table}

We compute $95\%$ confidence intervals using our neural estimator in two different ways to discuss the approximate distribution of the neural point estimator. In the first approach, the confidence interval endpoints are given by the $2.5\%$ and $97.5\%$ marginal percentiles of the estimated sampling distribution of $\widehat{\btheta}_r$ given by the bootstrap replicates: $(\widehat{\bq}_{r,0.025}, \widehat{\bq}_{r,0.975})$. In the second approach, we use a large sample Gaussian approximation to the distribution of $\widehat{\btheta}_r$ motivated in Section \ref{ss:inference}: $\widehat{\btheta}_r \pm 1.96 \widehat{\sf se}(\widehat{\btheta}_r)$, where $\widehat{\sf se}(\widehat{\btheta}_r)$ is the estimated standard error of $\widehat{\btheta}_r$ based on the standard deviation of the bootstrap replicates. We respectively refer to these two methods of computing confidence intervals as the percentile and Gaussian approximation methods. The coverage proportion (CP) of the $95\%$ confidence intervals averaged across the 500 Monte Carlo simulation replicates are reported in Table \ref{tab:small_sim_CI}. The Gaussian approximation appears to track the nominal level well for neural networks with smaller AVB. The percentile approach appears overly conservative, with coverage far above the nominal 95\% level. We hypothesize that this is because the parametric bootstrap fails to account for the uncertainty introduced by the estimation procedure in the sampling, so that the estimated sampling distribution of $\widehat{\btheta}_r$ is not necessarily symmetric around $\widehat{\btheta}_r$. 

\begin{table}[htb!]
\centering
\begin{tabular}{ccrrrrr}
 & parameter & \multicolumn{5}{c}{neural network} \\
 & & $r=1$ & $r=2$ & $r=3$ & $r=4$ & $r=5$\\
 \midrule
\multirow{2}{*}{percentile} & $\log(\phi^2)$ & 100 & 100 & 99.8 & 99.8 & 99.6  \\
& $\log(\tau^2)$ & 100 & 99.6 & 100 & 100 & 98.6 \\
\multirow{2}{*}{Gaussian approximation} & $\log(\phi^2)$ & 95.6 & 94.8 & 88.6 & 90.0 & 88.8 \\
& $\log(\tau^2)$ & 95.6 & 93.4 & 94.4 & 96.0 & 93.2
\end{tabular}
\caption{Coverage proportion (in \%) of 95\% confidence intervals for the non-distribution neural estimator in the Gaussian simulations. \label{tab:small_sim_CI}}
\end{table}

While the AVB appears a useful metric for selecting a good neural network, it requires fitting all 500 neural networks to the validation data for its evaluation. In the simulations of Section \ref{sex:simu}, we show that selecting the neural network with the smallest minimized validation loss, which is computed during training, also works well in practice due to the close relationship between AVB and the loss function.

\section{Distributed black-box inference} \label{s:dac}

\subsection{Problem setup}  \label{ss:setup}

We now consider the setting where $d$ is so large that estimation and inference for $\btheta$ based on $\mD$ is computationally intractable. In practice, if $Y(\mS)$ is a max-stable process then $d$ could be quite ``small'', whereas $d$ may be quite ``large'' when $Y(\mS)$ is, say, a Gaussian process. The point is that even for relatively inexpensive models like the Gaussian process there are values of $d$ that render inference on $\btheta$ intractable. In what follows we detail the main ingredients of our divide-and-conquer framework as well as a justification of the inference with our combined estimator.

\subsection{Local model fitting} \label{ss:local}

We propose to partition the spatial domain $\mD$ into $K$ disjoint regions $\mD_1, \ldots, \mD_K$ such that $\cup_{k=1}^K \mD_k=\mD$ and denote by $d_k$ the number of observation locations in $\mD_k$. To facilitate estimation of $\btheta$ in each subset $\mD_k$, we partition $\mD$ such that $d_k$ is relatively small, e.g. $d_k=20^2$ or $d_k=30^2$. While disjoint regions $\mD_1,\ldots, \mD_K$ are not technically required, overlapping regions may increase dependence between regions and incur numerical instability at the integration step; this is discussed further in Section \ref{ss:dac:integration}. The literature is rich with methods for choosing partitions for Gaussian and max-stable processes; see \cite{Heaton-etal-2019}, \cite{Hector-Reich} and \cite{Hector-Reich-Eloyan}.

For each $\btheta_t$, we simulate the outcome $Y_t(\mD_k)$, $t=1, \ldots, T$ on the spatial domain $\mD_k$ from $f \{ y(\mD_k); \btheta_t \}$. That is, we simulate data independently on each block of the partitioned spatial domain. Next, we train a neural network using $Y_t(\mD_k)$, $t=1, \ldots, T$ as the input and $\btheta_t$ as the output for each $k=1, \ldots, K$. In the special case with $d_k \equiv d_1$, then we need only simulate the outcome $Y_t(\mD_1)$, $t=1, \ldots, T$ on the spatial domain $\mD_1$ from $f \{ y(\mD_1); \btheta_t \}$ and train one neural network using $Y_t(\mD_1)$. While at first blush this may appear to substantially reduce the computing time of the approach, the simulation of $Y_t(\mD_k)$, $k=1, \ldots, K$, and the training of the $K$ neural networks can be performed in parallel across $K$ computing nodes, so that there is effectively no difference in total run time between $d_k \neq d_{k^\prime}$ and $d_k \equiv d_1$.

Denote by $a_k(\widehat{\bA}_k; \cdot)$ the network with input $\cdot$ and trained weights $\widehat{\bA}_k$. We use the trained neural network(s) to estimate the value of $\btheta_0$ for our observed data $Y(\mD_k)$, $k=1, \ldots, K$, yielding $K$ estimates $\widehat{\btheta}_k = a_k\{ \widehat{\bA}_k; Y(\mD_k)\}$, $k=1,\ldots, K$. We thus have computationally amortised estimates of $\btheta_0$ from each of the $K$ blocks. Section \ref{ss:dac:integration} describes the procedure for combining the estimates and inference over the $K$ blocks. 

\subsection{Neural network integration} \label{ss:dac:integration}

There are many potential approaches for combining estimates and inferences over the $K$ blocks. For example, the mean estimator $\widehat{\btheta}_m=K^{-1} \sum_{k=1}^K \widehat{\btheta}_k$ is a reasonable and computationally efficient point estimate of $\btheta_0$. Its variance, however, is susceptible to be inflated by any block estimates $\widehat{\btheta}_k$ with a large variance, as
\begin{align}
{\sf Var}(\widehat{\btheta}_m) &= \textstyle\sum_{k = 1}^K \bv_k + \textstyle\sum_{k,k^\prime = 1, ~k \neq k^\prime}^K {\sf Cov}(\widehat{\btheta}_k, \widehat{\btheta}_{k^\prime}), \label{e:var-mean}
\end{align}
with $\bv_k = {\sf Var} (\widehat{\btheta}_k)$. An alternative estimator that downweights block estimators with large variance is the inverse-variance weighted estimator, $\widehat{\btheta}_w = (\sum_{k=1}^K \bv_k^{-1} )^{-1} \sum_{k=1}^K \bv_k^{-1} \widehat{\btheta}_k$. While this weighted estimator will have a smaller variance than $\widehat{\btheta}_m$, one of the drawbacks is that it fails to account for the dependence ${\sf Cov}(\widehat{\btheta}_k, \widehat{\btheta}_{k^\prime})$ between $\widehat{\btheta}_k$ inherited from the dependence between $\mD_k$. Indeed, its variance is given by
\begin{align*}
\Bigl( \textstyle\sum_{k = 1}^K \bv_k^{-1} \Bigr)^{-1} + \Bigl( \textstyle\sum_{k = 1}^K \bv_k^{-1} \Bigr)^{-1} \Bigl\{  \textstyle\sum_{k,k^\prime = 1~k \neq k^\prime}^K \bv_k^{-1} {\sf Cov} (\widehat{\btheta}_k, \widehat{\btheta}_{k^\prime}) \bv_{k^\prime}^{-1} \Bigr\} \Bigl( \textstyle\sum_{k = 1}^K \bv_k^{-1} \Bigr)^{-1}.
\end{align*}
When ${\sf Cov} (\widehat{\btheta}_k, \widehat{\btheta}_{k^\prime})$ is positive-definite, as is often the case in spatial settings, the variance of $\widehat{\btheta}_w$ is strictly larger than $(\sum_{k=1}^K \bv_k^{-1})^{-1}$ in Loewner order. 

We propose an improved weighted estimator that accounts for this cross-covariance and thereby minimizes the variance of the resulting estimator. We proceed by first defining an estimating function for $\btheta$. Define $\bPsi_k(\btheta)= \widehat{\btheta}_k - \btheta$ so that solving $\bPsi_k(\btheta) =\bzero$ returns the estimator $\widehat{\btheta}_k$. 

Our proposed weighted estimator must account for and therefore estimate the dependence between $Y(\mD_k)$, $k=1, \ldots, K$. The major difference between our proposal here and previous work in, e.g. \cite{Hector-Reich} or \cite{Hector-Reich-Eloyan}, is that the outcome process is only observed once, and so bootstrapping is needed to quantify the dependence between $\widehat{\btheta}_k$. To this end, we sample independent replicates $Y_b(\mD_k)$, $b = 1, \ldots, B$ from $f\{ y (\mD_k) ;\widehat{\btheta}_m\}$ and generate bootstrap replicates in each block of the partitioned spatial domain using $\widehat{\btheta}_{kb} = a\{ \widehat{\bA}_k; Y_b(\mD_k)\}$, $b=1, \ldots, B$. This data generation and neural network fitting process can be run in parallel across up to $K$ computing nodes for maximal computational efficiency. The bootstrap replicates are only \emph{conditionally} independent across $k=1, \ldots, K$ given $\widehat{\bA}_k, Y_b(\mD_k)$: the distribution from which $Y_b(\mD_k)$ are sampled depends on $\widehat{\btheta}_m$, which is shared across the blocks and whose variance, given in equation \eqref{e:var-mean}, captures dependence between $\widehat{\btheta}_k$, $k=1, \ldots, K$. The bootstrap replicates $\widehat{\btheta}_{kb}$ are \emph{marginally} dependent across $k=1, \ldots, K$, with a dependence structure that captures the dependence between block estimators $\widehat{\btheta}_k$, $k=1, \ldots, K$. Intuitively, the covariance between the bootstrap replicates $\widehat{\btheta}_{kb}$ can be used to estimate the covariance between $\widehat{\btheta}_k$. To make this intuition formal, we construct a kernel estimating function for $\btheta$ in each block $\mD_k$ that depends on the bootstrap replicates $\widehat{\btheta}_{kb}$: define $\bpsi_{kb}(\btheta)=\widehat{\btheta}_{kb}-\btheta$. 

Define the stacking operation $\{\bb_k\}_{k=1}^K=(\bb^\top_1, \ldots, \bb_K)^\top \in \mathbb{R}^{\sum_{k=1}^K b_k}$ and $\{\bA_k\}_{k=1}^K=(\bB^\top_1, \ldots, \bB^\top_K)^\top \in \mathbb{R}^{\sum_{k=1}^K b_{1k} \times b_2}$ for $\bb_k \in \mathbb{R}^{b_k}$ and $\bB_k \in \mathbb{R}^{b_{1k}\times b_2}$, $k=1, \ldots, K$. Define the stacked kernel estimating and estimating functions
\begin{align*}
\bpsi_{b,all}(\btheta)&=\{ \bpsi_{kb}(\btheta)\}_{k=1}^K = \{ \widehat{\btheta}_{kb}-\btheta \}_{k=1}^K \in \mathbb{R}^{Kq},\quad b=1, \ldots, B,\\
\bPsi_{all}(\btheta)&= \{ \widehat{\btheta}_k-\btheta\}_{k=1}^K \in \mathbb{R}^{Kq},
\end{align*}
respectively. The stacked estimating function $\bPsi_{all}(\btheta)$ over-identifies $\btheta_0$: there are more estimating equations than there are dimensions on $\btheta$. \cite{Hansen}'s generalized method of moments (GMM) was designed for just such a setting. It minimizes a quadratic form of the over-identifying moment conditions:
\begin{align}
\widehat{\btheta}_{GMM}&=\arg \min \limits_{\btheta} \bPsi^\top_{all}(\btheta) \bW(\btheta) \bPsi_{all}(\btheta)=\arg \min \limits_{\btheta} \textstyle\sum_{k,k^\prime = 1}^K \bPsi^\top_k(\btheta) \left\{ \bW (\btheta) \right\}_{k,k'} \bPsi_{k'}(\btheta),
\label{e:GMM-quad}
\end{align}
where $\{ \bW (\btheta) \}_{k,k'}$ denotes the rows and columns of $\bW(\btheta)$ corresponding to subsets $\mD_k$ and $\mD_{k'}$ respectively, for any positive semi-definite weight matrix $\bW (\btheta) \in \mathbb{R}^{Kq \times Kq}$. \cite{Hansen} showed that the most efficient choice of $\bW(\btheta)$ is
\begin{align}\label{e:opt-weight}
\bW_{opt}(\btheta) &= 
\Bigl\{ \frac{1}{B-1} \sum \limits_{b=1}^B \bpsi_{b,all}(\btheta) \bpsi^\top_{b,all}(\btheta) \Bigr\}^{-1} \in \mathbb{R}^{Kq \times Kq}.
\end{align}
The matrix $\bW_{opt}^{-1}$ is a bootstrap estimator of $\bv(\btheta_0)={\sf Var}(\widehat{\btheta}_1^\top, \ldots, \widehat{\btheta}_K^\top)$ \citep{Carlstein, Lele, Sherman-Carlstein, Sherman, Heagerty-Lumley, Zhu-Morgan}. In fact, our approach has a unique advantage because $B$ can be made arbitrarily large subject to computational resource constraints, yielding an estimator $\bW_{opt}(\btheta)$ that can be made arbitrarily precise for $\bv(\btheta_0)$. \cite{Hansen} showed that this choice is \emph{Hansen optimal} in the sense that $\widehat{\btheta}_{opt}$, the value of $\btheta$ that minimizes $\bPsi_{all}^\top (\btheta) \bW_{opt}(\btheta) \bPsi_{all}(\btheta)$, has variance at least as small as any estimator given by equation \eqref{e:GMM-quad}.

Unfortunately, the computation of $\widehat{\btheta}_{opt}$ requires iterative minimization over all $K$ blocks simultaneously, which can be computationally expensive due to the need to recompute $\bW_{opt}(\btheta)$ at each iteration. Following \cite{Hector-Song-JASA} and \cite{Hector-Reich}, we define a one-step meta-estimator that is asymptotically equivalent to $\widehat{\btheta}_{opt}$ as $d_k \to \infty$. To this end, let $\bar{\btheta}_k=(1/B)\sum_{b=1}^B \widehat{\btheta}_{kb}$ and define $\widehat{\bpsi}_{b,all}=\{ \bpsi_{kb}(\bar{\btheta}_k) \}_{k=1}^K = \{\widehat{\btheta}_{kb} - \bar{\btheta}_k\}_{k=1}^K \in \mathbb{R}^{Kq}$, and
\begin{align*}
\widehat{\bW}_{opt}&= \Bigr( \frac{1}{B-1} \sum \limits_{b=1}^B \widehat{\bpsi}_{b,all} \widehat{\bpsi}_{b,all}^\top \Bigl)^{-1} \in \mathbb{R}^{Kq \times Kq}.
\end{align*}
Again, $\widehat{\bW}_{opt}^{-1}$ estimates $\bv(\btheta_0)$ and should be a more precise estimator as $B$ increases; see the justification in Section \ref{ss:inference}. The one-step estimator is defined as
\begin{align}
\widehat{\btheta}_c &= \Bigl\{ \textstyle\sum_{k,k^\prime = 1}^K (\widehat{\bW}_{opt})_{k,k'} \Bigr\}^{-1} \textstyle\sum_{k,k^\prime = 1}^K (\widehat{\bW}_{opt})_{k,k'} \widehat{\btheta}_{k'}.
\label{e:one-step}
\end{align}
The estimator $\widehat{\btheta}_c$ in equation \eqref{e:one-step} can be computed in an accelerated distributed data network without needing to reaccess individual data blocks nor perform any iterative optimization procedures. The estimation is amortised in the sense that no new training needs to occur when new data $Y(\mD^\prime)$ are collected so long as $\mD^\prime$ can be partitioned into blocks of sizes in $\{d_1, \ldots, d_K\}$. We discuss inference using $\widehat{\btheta}_c$ in Section \ref{ss:inference}.

\subsection{Uncertainty quantification}\label{ss:inference}

The bootstrap replicates obtained in Section \ref{ss:dac:integration} are much more computationally efficient to obtain than replicates on $\mD$ due to the reduced size of the spatial blocks, and thus facilitate computationally efficient point estimation. An important remaining question is whether the uncertainty of $\widehat{\btheta}_c$ in equation \eqref{e:one-step} can be quantified computationally efficiently as well. Recall that existing black-box approaches described in Section \ref{s:uncertainty-bb} simulate bootstrap replicates on the entire domain $\mD$ to estimate the sampling distribution of the black-box estimator. In contrast, a centerpiece of our proposal is an amortised inference, using $\widehat{\btheta}_c$, that only requires simulating the outcome on the blocks $\mD_k$. 

We have no general theoretical guarantees that $\widehat{\btheta}_k$, and therefore $\widehat{\btheta}_c$, is a consistent point estimator of $\btheta_0$ since $\widehat{\btheta}_k$ depends on the neural network structure and its estimated weights. \cite{gerber2021fast} provide empirical evidence that $\widehat{\btheta}_k - \widehat{\btheta}_k^\star = o_p(1)$ in a Gaussian process covariance model, where $\widehat{\btheta}_k^\star$ is the maximum likelihood estimator based on $Y(\mD_k)$. If this order of convergence holds, then 
\begin{align*}
\widehat{\btheta}_k - \btheta_0 &= o_p(1) + \widehat{\btheta}_k^\star - \btheta_0 = o_p(1),
\end{align*}
from which we can show that $\widehat{\btheta}_c \stackrel{p}{\to} \btheta_0$. To see why this is true, following similar arguments to \cite{Hector-Song-JASA}, define $\lambda(\btheta) =\sum_{k,k'=1}^K (\widehat{\bW}_{opt})_{k,k'} (\btheta - \widehat{\btheta}_{k'})$. Notice that $\lambda(\widehat{\btheta}_c) = \bzero$. Moreover, since $\widehat{\btheta}_k \stackrel{p}{\to} \btheta_0$,
\begin{align*}
\lambda (\btheta_0)&= \textstyle\sum_{k,k^\prime = 1}^K (\widehat{\bW}_{opt})_{k,k'} (\btheta_0 - \widehat{\btheta}_{k'}) = \textstyle\sum_{k,k^\prime = 1}^K \{ \bv_{k,k'} +o_p(1)\} o_p(1) = o_p(1).
\end{align*}
As $\nabla_{\btheta} \lambda(\btheta) = \sum_{k,k^\prime=1}^K (\widehat{\bW}_{opt})_{k,k^\prime}$ exists and is nonsingular, there exists some $\btheta^\dagger$ between $\btheta_0$ and $\widehat{\btheta}_c$ such that 
\begin{align*}
\lambda(\widehat{\btheta}_c) - \lambda(\btheta_0) &= \{ \nabla_{\btheta} \lambda(\btheta) \}_{\btheta=\btheta^\dagger} (\widehat{\btheta}_c - \btheta_0) = o_p(1).
\end{align*}
We conclude that $\widehat{\btheta}_c - \btheta_0 = o_p(1)$ and therefore $\widehat{\btheta}_c$ is a consistent estimator of $\btheta_0$. 

In fact, \cite{gerber2021fast} suggest that under some conditions $\widehat{\btheta}_k - \widehat{\btheta}_k^\star = o_p(d_k^{-1/2})$, and so
\begin{align*}
(\widehat{\btheta}_k-\btheta_0)_{k=1}^K &= \{ \widehat{\btheta}_k^\star - \btheta_0 + o_p(d_k^{-1/2})\}_{k=1}^K
\end{align*} 
is approximately Gaussian distributed with mean $\bzero$ and variance $\bv(\btheta_0)$ for large $d_k$. From this, we can show that $(\widehat{\btheta}_c-\btheta_0)$ is also approximately Gaussian distributed, also with mean $\bzero$ but with variance $\bj^{-1}(\btheta_0)=\{ \sum_{k,k^\prime=1}^K \bv^{-1}(\btheta_0) \}^{-1}$. Indeed, observe that
\begin{align*}
\bzero &= \lambda(\widehat{\btheta}_c) = \textstyle\sum_{k,k^\prime = 1}^K (\widehat{\bW}_{opt})_{k,k^\prime} (\widehat{\btheta}_c - \btheta_0 + \btheta_0 - \widehat{\btheta}_{k^\prime}).
\end{align*}
Rearranging, we obtain
\begin{align*}
\widehat{\btheta}_c - \btheta_0 &= \Bigl\{ \textstyle\sum_{k,k^\prime = 1}^K (\widehat{\bW}_{opt})_{k,k^\prime} \Bigr\}^{-1} \textstyle\sum_{k,k^\prime = 1}^K (\widehat{\bW}_{opt})_{k,k^\prime} (\widehat{\btheta}_{k^\prime} - \btheta_0)
\end{align*}
The right-hand side is approximately Gaussian distributed with mean $\bzero$ and variance $\bj^{-1}(\btheta_0)$ when $d_k$ are large, as stated above.

In order to use the proposed large sample distribution of $\widehat{\btheta}_c$ for inference about $\btheta_0$, we need an estimate of its large sample variance. As $\widehat{\bW}^{-1}_{opt}$ is an estimator of $\bv(\btheta_0)$ following the informal arguments in Section \ref{ss:dac:integration}, we can estimate $\bj(\btheta_0)$ with
\begin{align}
\widehat{\bJ}_{opt}&= \textstyle\sum_{k,k^\prime = 1}^K ( \widehat{\bW}_{opt} )_{k,k'}.
\label{e:J-hat}
\end{align}
This suggests the construction of large sample confidence intervals for $\btheta_0$ through
\begin{align}
\widehat{\btheta}_c \pm z_{\alpha/2} \Bigl[ {\sf diag} \Bigl\{ \textstyle\sum_{k,k^\prime = 1}^K ( \widehat{\bW}_{opt} )_{k,k'} \Bigr\}^{-1} \Bigr]^{1/2}.
\label{e:asy-ci}
\end{align}
We provide empirical evidence in Section \ref{sex:simu} that the distribution of $\widehat{\btheta}_c$ is Gaussian and centered at $\btheta_0$ with variance estimated by $\widehat{\bJ}_{opt}^{-1}$ when $a_k(\bA; \cdot)$ is sufficiently complex, and $d_k,B$ and $T$ are sufficiently large. 

\section{Simulation results} \label{sex:simu}
\subsection{Simulations with Gaussian processes}
\label{sub:gauss}

We investigate the performance of the weighted estimator $\widehat{\btheta}_c$ in equation \eqref{e:one-step}. The outcome $Y(\mD)$ is generated from a mean-zero Gaussian distribution with covariance function given in equation \eqref{e:gaussian-covariance}, with $\tau^2_0>0$ and $\phi^2_0>0$ the true variance and correlation parameters, respectively, and $\bs_j,\bs_{j^\prime} \in \mD$. We take $\mD=[1,d^{1/2}]^2$ to be a square gridded spatial domain of dimension $d^{1/2} \times d^{1/2}$. Our aim is to estimate and make inference on $\btheta_0=\{ \log(\tau^2_0), \log(\phi^2_0)\}$ when the number of locations $d$ is large.

In the first set of simulations, we fix $\tau^2_0=1$ and $\phi^2_0=3$ and vary $d \in \{60^2,90^2,120^2\}$. We partition the spatial domain $\mD$ into square blocks of $d_k=30^2$ locations each. Thus, $d=60^2,90^2,120^2$ respectively give $K=4, 9, 16$ blocks. Values of $\log(\phi^2_{t_1})$, $t_1=1, \ldots, T_1$, consist of the sequence from $\log(\phi^2_0)-0.5$ to $\log(\phi^2_0)+0.5$ (approximately $1.82$ to $4.95$) of length $T_1=150$, and values of $\log(\tau^2_{t_2})$, $t_2=1, \ldots, T_2$, consist of the sequence from $\log(\tau^2_0)-0.5$ to $\log(\tau^2_0)+0.5$ (approximately $0.607$ to $1.65$) of length $T_2=150$. Values $\btheta_t = \{ \log(\tau^2_{t_1}), \log(\phi^2_{t_2}) \}_{t_1,t_2=1}^{T_1,T_2}$, $t=1, \ldots, 22{,}500$, consist of all combinations of $\log(\tau^2_{t_1})$ and $\log(\phi^2_{t_2})$. For each $\btheta_t$, we simulate $Y_t(\mD_1)$ on $\mD_1=[1, 30]^2$. A validation sample is generated similarly to the training sample but with $T_1=T_2=70$. The training, validation and observed values of the outcome are transformed using $\mathbbm{1}(y >0)\log(y) -\mathbbm{1}(y \leq 0) \log(-y)$. The training and validation values of $\btheta$ are centered by the minimum training value and standardized by the range of the training values. The neural network architecture is described in Table \ref{tab:cnn-arch}. We train the network 500 times using the same architecture and training and validation data but 500 different seeds, and select the neural network with the smallest minimized validation loss. 

The neural network is used to fit the observed data $Y(\mD_k)$ in each block $k$ in 500 independent Monte Carlo simulation replicates. For each Monte Carlo replicate, the estimator in \eqref{e:one-step} is computed using $B=5{,}000$ bootstrap replicates. Computations in blocks are parallelized across $K$ CPUs. In Table \ref{tab:gaussian-I-results}, we report root mean squared error (RMSE), empirical standard error (ESE), average asymptotic standard error (ASE) using the formula in equation \eqref{e:J-hat}, and average 95\% confidence interval coverage (CP) using the formula in equation \eqref{e:asy-ci} averaged across the 500 simulations. When  $d=60^2$, we also compare our estimator's MSE and ESE to that of the maximum likelihood estimator (MLE) fit on the whole domain and the Vecchia approximation \citep{vecchia1988estimation} using the $d_k=30^2$ nearest neighbours implemented using the \texttt{R} package \texttt{GpGp} \citep{GpGp2, GpGp1}. 

\begin{table}[htb!]
\centering
\begin{tabular}{cccccc} 
$d$ & parameter & RMSE$\times 100$ & ESE$\times 100$ & ASE$\times 100$ & CP (\%)\\
\midrule
\multirow{2}{*}{$60^2$} & $\log(\phi^2)$ & 7.81 (9.12,9.45) & 7.80 (9.11,9.43) & 7.76 & 93.4 \\ 
& $\log(\tau^2)$ & 7.55 (8.58,8.69) & 7.54 (8.57,8.55) & 7.09 & 92.4 \\  
\multirow{2}{*}{$90^2$} & $\log(\phi^2)$ & 5.70 & 5.67 & 5.19 & 92.6 \\  
& $\log(\tau^2)$ & 5.23 & 5.20 & 4.74 & 92.2 \\ 
\multirow{2}{*}{$120^2$} & $\log(\phi^2)$ & 4.12 & 4.05 & 3.89 & 94.2 \\ 
& $\log(\tau^2)$ & 3.77 & 3.71 & 3.56 & 94.2 \\ 
\end{tabular}
\caption{Simulation metrics for the distributed neural estimator in the first set of Gaussian simulations, with simulation metric of MLE and Vecchia approximation, respectively, in parentheses.}
\label{tab:gaussian-I-results}
\end{table}

The RMSE and ESE are approximately equal, suggesting that the bias of $\widehat{\btheta}_c$ is negligible. Further, the ESE and ASE are approximately equal, supporting the use of the estimator $\sum_{k,k^\prime=1}^K (\widehat{\bW}_{opt})_{k,k^\prime}$ in equation \eqref{e:J-hat} for the inverse variance of $\widehat{\btheta}_c$. Finally, 95\% confidence intervals computed using equation \eqref{e:asy-ci} reach their nominal levels, suggesting that the Gaussian approximation to the distribution of $\widehat{\btheta}_c$ works well for inference when $d_k$ are large. The MLE isslightly less efficient than the combined estimator $\widehat{\btheta}_c$, potentially due to the use of the bounded training set $\btheta_t$. Further, the Vecchia is slightly less efficient than the MLE, suggesting that $30^2$ neighbours are insufficient to capture the full range of spatially dependent neighbours.

Mean elapsed time (standard error) in seconds for our estimator is $13$ ($4.84$), $14.4$ ($5.67$), $15.5$ ($7.99$) for $=60^2,90^2,120^2$ respectively. The mean elapsed time remains relatively constant as $d$ increases: this makes sense, as the entire procedure only requires simulation over the same $\mD_1=[1, 30]^2$ domain. In comparison, the mean elapsed time (standard error) in seconds of the MLE and Vecchia approximation when $d=60^2$ is $960$ ($264$) and $1690$ ($411$) respectively. The distributed approach is thus much faster than both the MLE and Vecchia approximation, highlighting the considerable computational advantage of our distributed approach.

In the second set of simulations, we fix $d=90^2$ and vary $\tau^2_0 \in \{0.5,1,1.5,2\}$ and $\phi^2_0 \in \{2,3,4\}$ for a total of $11$ combinations: $(\phi^2_0,\tau^2_0) \in \{(2,0.5), (3,0.5), (4,0.5), (2, 1), (4,1), (2,1.5), \allowbreak (3,1.5), (4,1.5), (2,2), (3,2), (4,2)\}$. We partition the spatial domain $\mD$ into square blocks of $d_k=30^2$ locations each, giving $K=9$ blocks. Values of $\log(\phi^2_{t_1})$, $t_1=1, \ldots, T_1$, consist of the sequence from $\log(\phi^2_0)-0.5$ to $\log(\phi^2_0)+0.5$ of length $T_1=150$, and values of $\log(\tau^2_{t_2})$, $t_2=1, \ldots, T_2$, consist of the sequence from $\log(\tau^2_0)-0.5$ to $\log(\tau^2_0)+0.5$ of length $T_2=150$. Values $\btheta_t = \{ \log(\tau^2_{t_1}), \log(\phi^2_{t_2}) \}_{t_1,t_2=1}^{T_1,T_2}$, $t=1, \ldots, 22{,}500$, consist of all combinations of $\log(\tau^2_{t_1})$ and $\log(\phi^2_{t_2})$. Training values of the outcomes are generated as in the first set of simulations on $\mD_1=[1,30]^2$. A validation sample is generated similarly to the training sample but with $T_1=T_2=70$. Transformation of training, validation and observed values of the outcome and standardization of training, validation and testing values of $\btheta_t$ are as in the first set of simulations. The neural network architecture and training is performed as in the first set of simulations. 

The neural network with smallest minimized validation loss out of 500 trained networks is used to fit the observed data $Y(\mD_k)$ in each block $k$ in 500 independent Monte Carlo simulation replicates. For each Monte Carlo replicate, the estimator in \eqref{e:one-step} is computed using $B=5{,}000$ bootstrap replicates. Computations in blocks are parallelized across $K=9$ CPUs. We report RMSE, ESE, ASE and CP averaged across the 500 simulations in Table \ref{tab:gaussian-II-results}. Across all settings, ESE and ASE are approximately equal, again suggesting the suitability of equation \eqref{e:J-hat} to estimate the inverse variance of $\widehat{\btheta}_c$. The RMSE is close to the ESE and ASE in most settings, but grows slightly larger when the spatial dependence and variance are weaker ($\phi^2=2$, $\tau^2=0.5$) and when the spatial dependence and variance are stronger ($\phi^2=4$, $\tau^2=2$). This suggests that the CNN exhibits a small amount of bias in these more difficult settings. Finally, the CP seems slightly lower than the nominal coverage rate, presumably due to the non-zero bias of the CNN. The CP reaches the nominal level in almost all settings.

\begin{table}[htb!]
\centering
\begin{tabular}{ccccccc} 
$\phi^2$ & $\tau^2$ & parameter & RMSE$\times 100$ & ESE$\times 100$ & ASE$\times 100$ & CP (\%)\\
\midrule
\multirow{2}{*}{2} & \multirow{2}{*}{0.5} & $\log(\phi^2)$ & 7.09 & 6.42 & 6.09 & 90.4 \\ 
& & $\log(\tau^2)$ & 5.73 & 5.66 & 5.35 & 93.0 \\ 
\multirow{2}{*}{3} & \multirow{2}{*}{0.5} & $\log(\phi^2)$ & 5.96 & 5.92 & 5.51 & 93.4 \\ 
& & $\log(\tau^2)$ & 5.10 & 4.88 & 4.76 & 93.0 \\ 
\multirow{2}{*}{4} & \multirow{2}{*}{0.5} & $\log(\phi^2)$ & 5.82 & 5.83 & 5.14 & 91.6 \\ 
& & $\log(\tau^2)$ & 5.45 & 5.16 & 4.88 & 91.8 \\ 
\multirow{2}{*}{2} & \multirow{2}{*}{1} & $\log(\phi^2)$ & 5.92 & 5.89 & 5.65 & 94.4 \\ 
& & $\log(\tau^2)$ & 5.11 & 5.12 & 4.97 & 93.8 \\ 
\multirow{2}{*}{4} & \multirow{2}{*}{1} & $\log(\phi^2)$ & 5.30 & 5.26 & 5.00 & 92.6 \\ 
& & $\log(\tau^2)$ & 5.83 & 5.75 & 5.20 & 91.4 \\ 
\multirow{2}{*}{2} & \multirow{2}{*}{1.5} & $\log(\phi^2)$ & 5.44 & 5.41 & 5.44 & 94.4 \\ 
& & $\log(\tau^2)$ & 5.27 & 5.27 & 4.95 & 93.8 \\ 
\multirow{2}{*}{3} & \multirow{2}{*}{1.5} & $\log(\phi^2)$ & 5.21 & 5.11 & 5.16 & 94.0 \\ 
& & $\log(\tau^2)$ & 5.39 & 5.40 & 5.33 & 94.0 \\ 
\multirow{2}{*}{4} & \multirow{2}{*}{1.5} & $\log(\phi^2)$ & 5.90 & 5.75 & 5.37 & 90.8 \\ 
& & $\log(\tau^2)$ & 5.37 & 5.37 & 5.04 & 93.8 \\ 
\multirow{2}{*}{2} & \multirow{2}{*}{2} & $\log(\phi^2)$ & 5.99 & 5.63 & 5.36 & 92.2 \\ 
& & $\log(\tau^2)$ & 4.81 & 4.81 & 4.65 & 94.4 \\ 
\multirow{2}{*}{3} & \multirow{2}{*}{2} & $\log(\phi^2)$ & 5.25 & 5.22 & 5.11 & 93.0 \\ 
& & $\log(\tau^2)$ & 5.53 & 5.11 & 4.79 & 91.2 \\ 
\multirow{2}{*}{4} & \multirow{2}{*}{2} & $\log(\phi^2)$ & 5.57 & 5.58 & 5.38 & 93.8 \\ 
& & $\log(\tau^2)$ & 5.9 & 5.84 & 5.16 & 90.0 \\ 
\end{tabular}
\caption{Simulation metrics for the distributed neural estimator in the second set of Gaussian simulations.}
\label{tab:gaussian-II-results}
\end{table}

We report mean elapsed time (standard error) in seconds for our distributed approach in Table \ref{tab:gaussian-II-time}. Again, the mean elapsed time remains relatively constant as $d$ increases, with minor variations due to differences in computing nodes. These results suggest that elapsed time is insensitive both to the total dimension $d$ of the domain and to the difficulty of the estimation problem stemming from weak or strong spatial dependence ($\phi_0^2, \tau_0^2$).

\begin{table}[htb!]
\centering
\begin{tabular}{c|cccc} 
& \multicolumn{4}{c}{$\tau^2$} \\
$\phi^2$ & 0.5 & 1 & 1.5 & 2 \\
\midrule
2 & 21.9 (9.21) & 15.3 (5.70) & 14.1 (2.88) & 13.6 (3.94)\\ 
3 & 14.9 (4.90) &  & 13.8 (2.91) & 14.8 (7.19)\\ 
4 & 19.8 (13.0) & 21.4 (8.47) & 14.7 (5.55) & 16.8 (6.85)
\end{tabular}
\caption{Mean elapsed time (standard error) in seconds for our distributed approach in the second set of Gaussian simulations. \label{tab:gaussian-II-time}}
\end{table}

\subsection{Simulations with Brown-Resnick processes}
\label{sub:BR}

We now assess the performance of $\widehat{\btheta}_c$ with the Brown-Resnick model, which is commonly used to describe spatial extremes for max-stable processes \citep{kabluchko2009stationary}. These processes have an intractable likelihood, making maximum likelihood or Bayesian inference effectively impossible, even in small dimensions. While previous work has shown the effectiveness of black box approaches in statistical inference with the Brown-Resnick model \citep[see][and references therein]{lenzi2021neural}, they have only been considered in moderate dimensions, with at most $d=30^2$ spatial locations. 

The spatial dependence in the Brown-Resnick model is characterized by a zero-mean Gaussian process with semivariogram $\gamma(\bh) = (\lVert\bh\rVert/\lambda)^\nu$, where $\bh$ is the spatial distance, $\lambda > 0$ is the range,  $\nu \in (0, 2]$ is the smoothness and with margins assumed to be unit Fr\'echet.  To facilitate estimation, we estimate parameters on the transformed scale \citep{lenzi2021neural}: $ \btheta= (\theta_1, \theta_2) = [\log(\lambda), \log \{ \nu/(2 - \nu) \} ]$.

As in the Gaussian example, we fix $\mD=[1,d^{1/2}]^2$ to be a square gridded spatial domain of dimension $d^{1/2} \times d^{1/2}$. 
In the first set of simulations, we set $\lambda_0=1$, $\nu_0=1$ and vary $d \in \{20^2,30^2,40^2,50^2\}$, and in the second set of simulations we fix $d=20^2$ and vary $\lambda \in \{0.5,1,1.5\},\nu \in \{0.5,1,1.5\}$. 
We partition the spatial domain $\mD$ into square blocks of $d_k=10^2=100$ locations each such that we have $K = 4, 9, 16, 25$ blocks. Values of $\log(\lambda_{t_1})$, $t_1=1, \ldots, T_1$, consist of the sequence from $-0.7$ to $0.7$ of length $T_1=70$, and values of $\log\{\nu_{t_2}/(2-\nu_{t_2})\}$, $t_2=1, \ldots, T_2$, consist of the sequence from $0$ to $2$ of length $T_2=70$. Values $\btheta_t$, $t=1, \ldots, 4{,}900$, consist of all combinations of $\log(\lambda_{t_1})$ and $\log\{\nu_{t_2}/(2-\nu_{t_2})\}$. Next, for each pair of training parameter values, we simulate data $Y_t(\mD_1)$ on $\mD_1=[1,10]^2$ from the Brown-Resnick model. A validation sample is generated similarly to the training sample but with $T_1=T_2=16$. The training, validation and observed values of the outcome are transformed using $\log(y)$. The training and validation values of $\btheta$ are centered by the minimum training value and standardized by the range of the training values. The neural network architecture is described in Table \ref{tab:cnn-arch}. We train the network 500 times using the same architecture and training and validation data but 500 different seeds, and select the neural network with the smallest minimized validation loss. 

The neural network is used to fit the observed data $Y(\mD_k)$ in each block $k$ in 500 independent Monte Carlo simulation replicates. For each Monte Carlo replicate, the estimator in \eqref{e:one-step} is computed using $B=10{,}000$ bootstrap replicates. Computations in blocks are parallelized across $K$ CPUs. As a comparison, we also obtain the tapered composite likelihood estimator \citep{padoan2010likelihood} using observation pairs within a certain cutoff distance. The cutoff distance is specified such that that the resulting pairwise likelihood (PL) function contains only the $100$ observation pairs with the smallest distance. The PL is fit to each $Y(\mD)$ independently using the \texttt{SpatialExtremes} \texttt{R} package \citep{ribatet2008spatialextremes}, using the respective true parameter values as initial values in the optimization.
Table \ref{tab:BR-I-results} displays the RMSE, ESE, ASE and CP across the 500 simulations using our approach and the PL when $\theta_1=\theta_2=1$ and $d \in \{20^2,30^2,40^2,50^2\}$, and Table \ref{tab:BR-II-results} displays these same metrics when $d=20^2$ and $\lambda \in \{0.5,1,1.5\},\nu \in \{0.5,1,1.5\}$.

From Table \ref{tab:BR-I-results}, we see that the RMSE, ESE and ASE values tend to decrease as $d$ increases, suggesting that the accuracy or the combined estimator in larger areas improves. 
For all values of $d$, the ESE and ASE are approximately equal, suggesting that equation \eqref{e:J-hat} is also appropriate for estimating the inverse variance of $\widehat{\btheta}_c$.
The RMSE and ESE estimates from the pairwise likelihood are substantially larger for all settings, even when the optimization was initialized with the true value.  
The CP is close to the nominal coverage rate for both parameters and all settings, with only slightly undercoverage in a few cases.
These metrics indicate that our estimator has a low bias independent of the window size when $\theta_1=\theta_2=1$. When comparing the performance of our estimator for different combinations of true parameter values, we notice slightly larger RMSE, ESE, and ESE values for $\theta_2$, especially for low values of this parameter, when the field is less smooth.
The CP is below the nominal coverage, especially for larger values of $\lambda$,  indicating that estimated confidence intervals might be too narrow or slightly biased when the spatial dependence is relatively strong. 
The pairwise likelihood results, shown in parenthesis, are once again significantly worse in terms of both RMSE and ESE.

\begin{table}[h]
\centering
\begin{tabular}{cccccc} 
$d$ & Parameter & RMSE$\times 100$ & ESE$\times 100$ & ASE$\times 100$ & CP (\%) \\
\midrule
\multirow{2}{*}{$20^2$} & $\theta_1$ & 10.2 (83.2) & 10.1 (80.8) & 10.1 & 94.4  \\
&  $\theta_2$ & 13.6 (109) & 13.5 (106) & 12.7 & 94.4   \\
\multirow{2}{*}{$30^2$} & $\theta_1$ & 6.50 (115) & 6.43 (112) & 6.75 & 95.2  \\
&  $\theta_2$ & 8.64 (169) & 8.65 (160) & 8.46 & 94.6  \\
\multirow{2}{*}{$40^2$} & $\theta_1$ & 5.30 (91.1) & 5.26 (90.1) & 5.06 & 94.3 \\
&  $\theta_2$ & 6.57 (166) & 6.55 (156) & 6.34 & 94.3  \\ 
\multirow{2}{*}{$50^2$} & $\theta_1$ &  4.23 (88.6) & 4.17 (88.7) & 4.05 & 95.0  \\
&  $\theta_2$ & 5.19 (206) & 5.19 (180) & 5.07 & 93.8    \\ 
\end{tabular}
\caption{Simulation metrics for the distributed neural estimator in the first set of Brown-Resnick simulations with $\theta_1 = \theta_2 = 1$ with simulation metric of PL in parentheses.}
\label{tab:BR-I-results}
\end{table}

\begin{table}[h]
\centering
\begin{tabular}{ccccccc} 
$\lambda$ & $\nu$ & parameter & RMSE$\times 100$ & ESE$\times 100$ & ASE$\times 100$ & CP (\%)\\
\midrule
\multirow{2}{*}{0.5} & \multirow{2}{*}{0.5} & $\theta_1$ & 13.9 (143) & 11.9 (141) & 12.2 & 91.8  \\ 
& & $\theta_2$ & 14.1 (99.1) & 13.8 (96.3) & 12.5 & 90.0 \\ 
\multirow{2}{*}{1} & \multirow{2}{*}{0.5} & $\theta_1$ & 12.5 (122) & 12.5 (119) & 11.9 & 93.0\\ 
& & $\theta_2$ & 13.5 (106) & 13.3 (105) & 11.8 & 91.6 \\ 
\multirow{2}{*}{1.5} & \multirow{2}{*}{0.5} & $\theta_1$ & 13.2 (120) & 12.1 (117) & 10.7 & 87.8  \\ 
& & $\theta_2$ & 12.0 (120) & 11.9 (120) & 11.3 & 94.4 \\ 
\multirow{2}{*}{0.5} & \multirow{2}{*}{1} & $\theta_1$ & 11.0 (80.5) & 10.8 (79.7) & 10.3 & 91.8 \\ 
& & $\theta_2$ & 10.3 (109) & 7.58 (107) & 7.99 & 89.5 \\ 
\multirow{2}{*}{1.5} & \multirow{2}{*}{1} & $\theta_1$ & 11.5 (78.7) & 10.7 (76.6) & 10.3 & 90.4 \\ 
& & $\theta_2$ & 14.5 (119) & 12.9 (117) & 12.3 & 89.4 \\ 
\multirow{2}{*}{0.5} & \multirow{2}{*}{1.5} & $\theta_1$ & 9.29 (92.0) & 8.75 (87.9) & 8.99 & 93.0 \\ 
& & $\theta_2$ & 10.9 (198) & 10.6 (189) & 10.6 & 94.0 \\ 
\multirow{2}{*}{1} & \multirow{2}{*}{1.5} & $\theta_1$ & 8.95 (58.5) & 8.57 (57.3) & 8.09 & 91.3 \\ 
& & $\theta_2$ & 11.7 (185) & 11.5 (180) & 11.6 & 96.0 \\ 
\multirow{2}{*}{1.5} & \multirow{2}{*}{1.5} & $\theta_1$ & 8.69 (53.5) & 7.75 (52.8) & 7.71 & 91.1 \\ 
& & $\theta_2$ & 12.0 (188) & 11.0 (186) & 11.2 & 91.8 \\ 
\end{tabular}
\caption{Simulation metrics for the distributed neural estimator in the first set of Brown-Resnick simulations with $d=20^2$, with simulation metric of PL in parentheses.}
\label{tab:BR-II-results}
\end{table}

We report mean elapsed time (standard error) in seconds for our distributed approach as well as the pairwise likelihood in Table \ref{tab:BR-II-time}. The mean elapsed time is about $20\%$ faster and less variable for smaller values of $\nu$, namely smoother fields. 
The pairwise likelihood is considerably slower than our distributed approach across all parameter values.

\begin{table}[h]
\centering
\begin{tabular}{cccc} 
$\lambda$ & $\nu$ & distributed approach & PL\\
\midrule
0.5 & 0.5 & 222 (17.3)  & 415 (25.0)  \\ 
1 & 0.5 & 229 (22.7) & 366 (36.2)  \\ 
1.5 & 0.5 & 243 (18.9) & 376 (34.5) \\ 
0.5 & 1 & 281 (18.6) & 347 (16.0)  \\ 
1.5 & 1 & 262 (15.7) & 282 (15.3)  \\ 
0.5 & 1.5 & 280 (53.8) & 376 (23.1)  \\ 
1 & 1.5 & 278 (38.6) & 324 (18.2) \\
1.5 & 1.5 & 280 (46.4) & 337 (21.4) \\
\end{tabular}
\caption{Mean elapsed time (standard error) in seconds for the Brown-Resnick model for our distributed approach and PL estimator. \label{tab:BR-II-time}}
\end{table}

\section{Data analysis}\label{s:data}

\subsection{Land-surface temperature dataset} \label{s:data-reanalysis}
 
We use our distributed method to investigate the spatial extent of regions simultaneously affected by extreme temperature events using standardized annual reanalysis temperature maxima generated by the North American Land Data Assimilation System (NLDAS-2) \citep[see][for a description of the dataset]{mitchell2004multi}.
The data are in a 0.125-degree grid spacing and range from 01 January 1979 to 31 December 2023 with a monthly temporal resolution. We select a window of $180 \times 180$ spatial locations in the center of the United States, yielding $d = 180^2 = 32{,}400$ spatially dependent monthly temperature maximum observations from 01 January 1979 to 31 December 2023. This is an extremely large spatial domain where, in principle, it would be infeasible to fit a max stable process, even with approximations. To the best of our knowledge, ours is the first attempt to jointly model the spatial extremal dependence of such a large area.

\begin{figure}[h!]
\centering
\includegraphics[trim={0 8em 0 8em}, clip, width=0.8\textwidth]{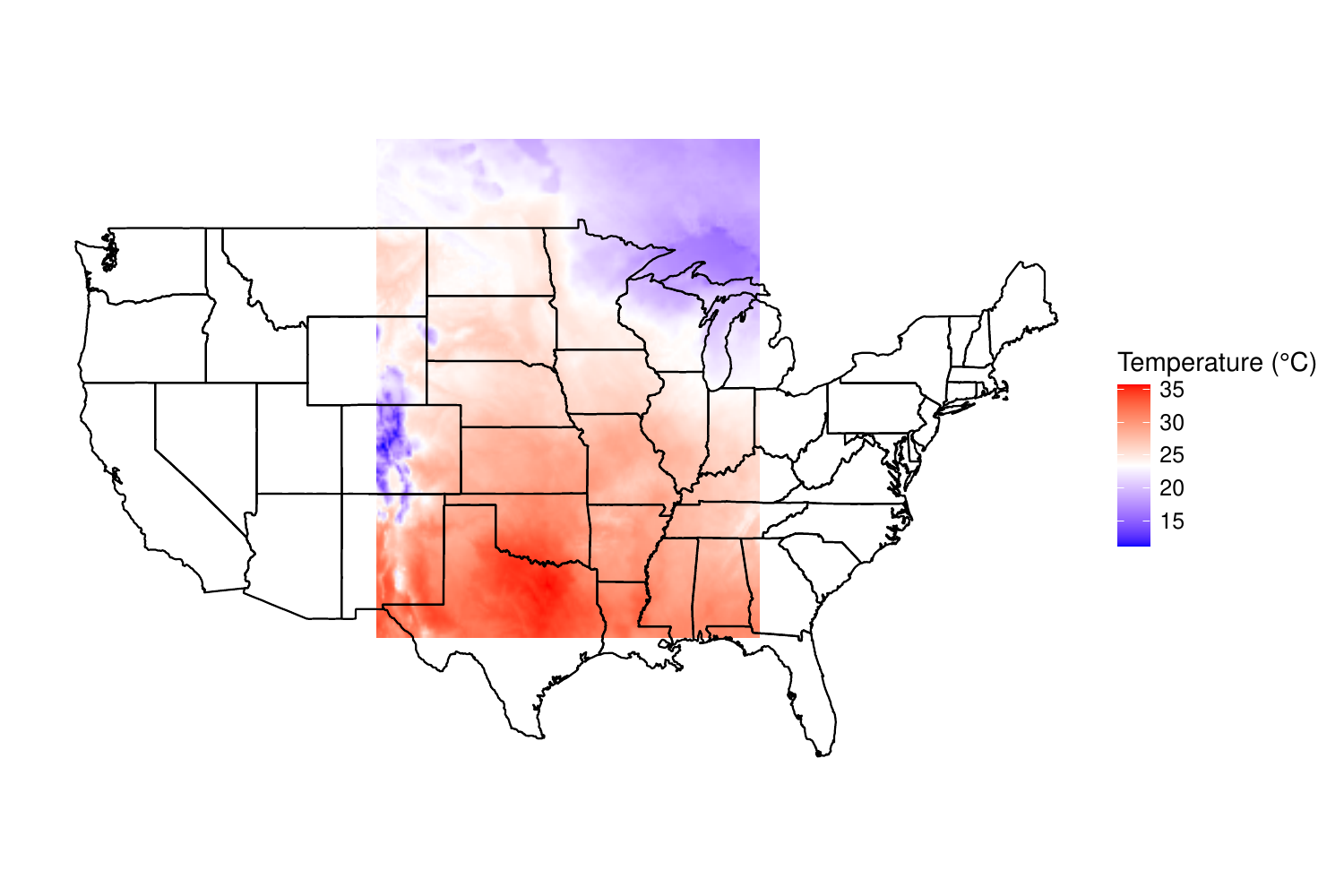}
\caption{Temperature maxima from the NLDAS-2 dataset for 2023.}
\label{f:NLDAS_map}
\end{figure}

Figure \ref{f:NLDAS_map} shows an example of the maximum temperature in degrees Celsius for 2023 over the selected locations. This figure shows that the annual temperature maxima are highly spatially correlated, with generally warmer temperatures in the southwest part of the United States and a gradual change to colder temperatures going North.

Before modeling the spatial extremal dependence, we first detrend the marginal distributions of the temperature measurements to account for the effect of global warming and standardize them to a common scale. To estimate spatiotemporal trends, we fit a spatiotemporal Gaussian model with a common yearly trend to all annual temperature maxima. Then, we use the residual from the fitted model to extract annual temperature maxima and fit a generalized extreme-value (GEV) distribution to each spatial location separately. We then obtain marginals that are unit Fr'{e}chet by transforming the data using the estimated parameters from the marginal GEV fits, which are then modeled with a Brown-Resnich max stable process.

As in Section \ref{sub:BR}, the dependence parameter vector to be estimated is $\btheta= [\mbox{log}(\lambda), \allowbreak \mbox{log}\{(\nu)/(2-\nu)\}]^{\top}$. The first step consists of dividing the spatial domain into smaller regions and obtaining local estimates for each block separately. We partition the $180 \times 180$ grid into $K=9^2 = 81$ blocks of size $d_k=20 \times 20$, which gives a good compromise of having enough data to learn the dependencies and being small enough that it can be simulated relatively quickly. 
We first fit a Brown–Resnick model in each block $\mD_k$ using the weighted pairwise likelihood, with the likelihood function containing at most second-order neighbors and binary weights. 
We find pairwise likelihood optimization to be highly sensitive to the initial values. 
Therefore, we give the optimizer multiple starting values of the range parameter $\lambda$ ranging from 1 to 20 (roughly the size of the domain) while keeping the smoothness equal to 1. 
We select the value of $\lambda$ with the highest pairwise likelihood and use it as the starting point to re-run the full optimization. The resulting estimates of $\lambda$ are roughly equal to 17, which we use as an initial guess to generate training data for the neural network.

We generate $70^2$ training datasets on a regular grid with values of $\theta_1$ ranging from $\mbox{log}(17)-2$ to $\mbox{log}(17)+2$ and $\theta_2$ equally spaced over the entire bounded domain. We train a neural network with the architecture described in Table \ref{tab:cnn-arch}  using standardized log-transformed annual maxima simulated from a Brown–Resnick max-stable model. The fitted neural network is then used to obtain parameter estimates in each block and each year separately. 

\begin{figure}[h!]
\centering	    
\includegraphics[width=0.92\textwidth]{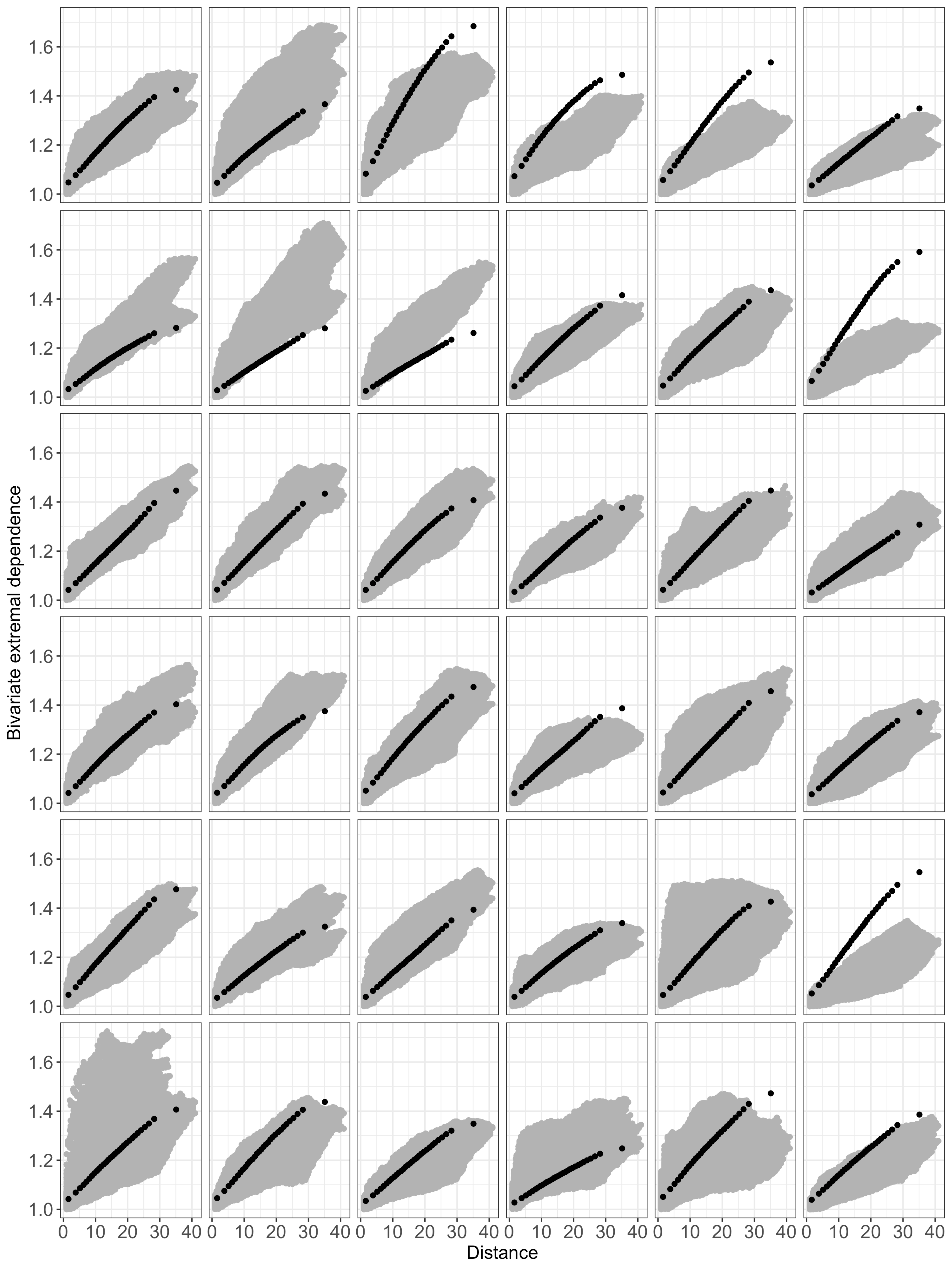}
\caption{Empirical bivariate extremal coefficients from the NLDAS-2 data (grey dots) plotted as a function of the Euclidean distance, and their model-based counterpart (black dots). Each panel corresponds to a block of size $d=30^2$.}
\label{f:extcoeff_nldas}
\end{figure}

We first examine the neural network fits in each block before combining them and computing $\widehat{\btheta}_c$. To visualize the spatial extremal dependence of the empirical versus fitted values, we use the so-called extremal coefficient, which gives a measure of extremal dependence between two stationary max-stable random fields. Coefficient values close to 1 correspond to perfect dependence, and 2 corresponds to independence. Due to the computational complexity of computing extremal coefficients for every pair of locations, we are not able to evaluate these estimates for large areas, and so we focus on subsets of the domain $\mD$ of size $30^2$, slightly larger than the blocks used to train the neural network. The empirical values of the extremal coefficients are calculated using the 45 years of available data as replicates in each of these $30^2$ subsets and are shown in Figure \ref{f:extcoeff_nldas} as grey dots. 
The black dots are binned extremal coefficients computed from simulating 45 replicates from a Brown-Resnick process of size $30^2$ with parameters set to the neural network estimates in each block. Our proposed method appears to capture the spatial extremal dependence well for most blocks and distances. In very few cases, especially at larger distances, the model can either over or underestimate extremal dependence, showing no particular bias.

\begin{figure}[h!]
\centering	    
\includegraphics[width=0.92\textwidth]{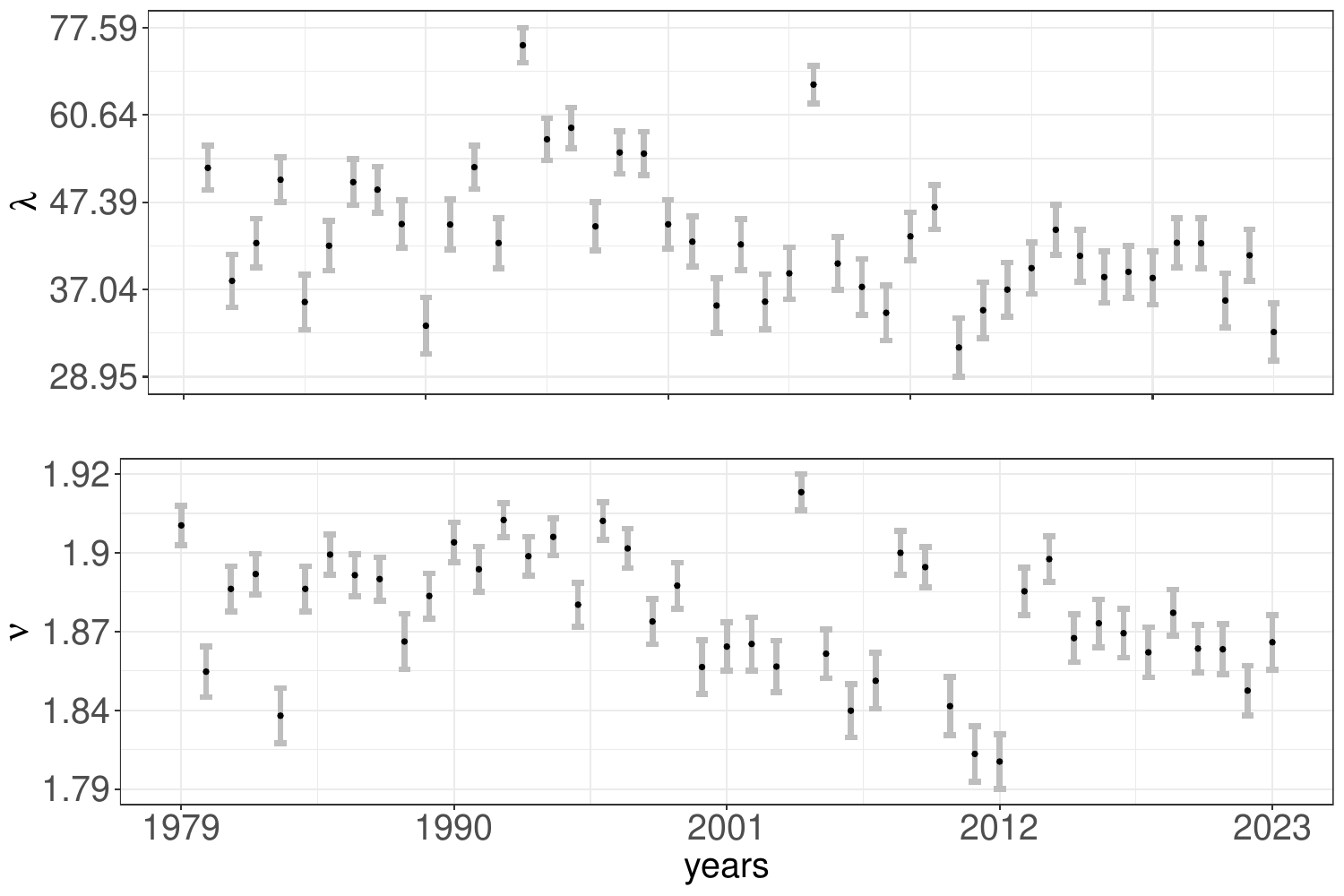}
\caption{$95\%$ confidence intervals of the combined weighted estimator $\widehat{\btheta}_c$ of the range parameter (top) and smoothness (bottom) on the transformed scales (with numbers on the axes on the untransformed scales) based on the NLDAS-2 dataset. The confidence intervals are calculated and displayed separately for the 45 years of data. Black dots indicate the combined weighted estimator.}
\label{f:ci95_nldas}
\end{figure}

We then compute the combined weighted estimator $\widehat{\btheta}_c$ using $B = 5{,}000$ bootstrap replicates. Figure \ref{f:ci95_nldas} shows $95\%$ confidence intervals of $\widehat{\btheta}_c$ calculated using the formula in equation \eqref{e:asy-ci} for the range and smoothness on the transformed scales (with numbers on the axes on the untransformed scales) with a separate bar for each year. The interval widths are similar for most years, and we observe a temporal trend, with estimated values decreasing after 1990 and increasing after 2012. 
As expected, the intervals always contain the simple average of the estimates in each block (black dots). 

\subsection{United States temperature dataset}  
\label{s:data-observed}

While reanalysis datasets are spatially and temporally consistent, they may struggle to accurately represent extremes, as they rely on the assimilated data and model physics. Here, we apply our distributed approach to extreme temperature data from the NOAA Monthly U.S. Climate Gridded Dataset (NClimGrid) \citep{vose2014noaa}.
The dataset is freely available from 01 January 1895 to the present and provides gridded in situ measurements to the public for, e.g. calculations of regional spatiotemporal climate trends. The data are monthly values in a $5\times5$ latitude/longitude grid for the Continental United States.

We compute yearly temperature maxima for the 129 years for a spatial region in the center of the United States of size $d=180^2=32{,}400$, the same number of locations considered in the NLDAS-2 data analysis (see Section \ref{s:data-reanalysis}). 
We fit a GEV distribution for each spatial location separately, which we then use to transform annual maxima to a common unit Fr\'echet scale with the probability integral transform. We model the spatial dependence structure of the standardized log-transformed annual maxima with a Brown–Resnick max-stable model. For this purpose, we use the trained CNN from Section \ref{s:data-reanalysis}, such that we need only plug in the new data to obtain estimates for each block and year. This reinforces the power of our method's amortization, where the neural network only needs to be trained once, and can easily be applied to model different datasets of possibly different sizes, as one only needs to adjust for the number of blocks. 

\begin{figure}[h!]
\centering	    
\includegraphics[width=0.92\textwidth]{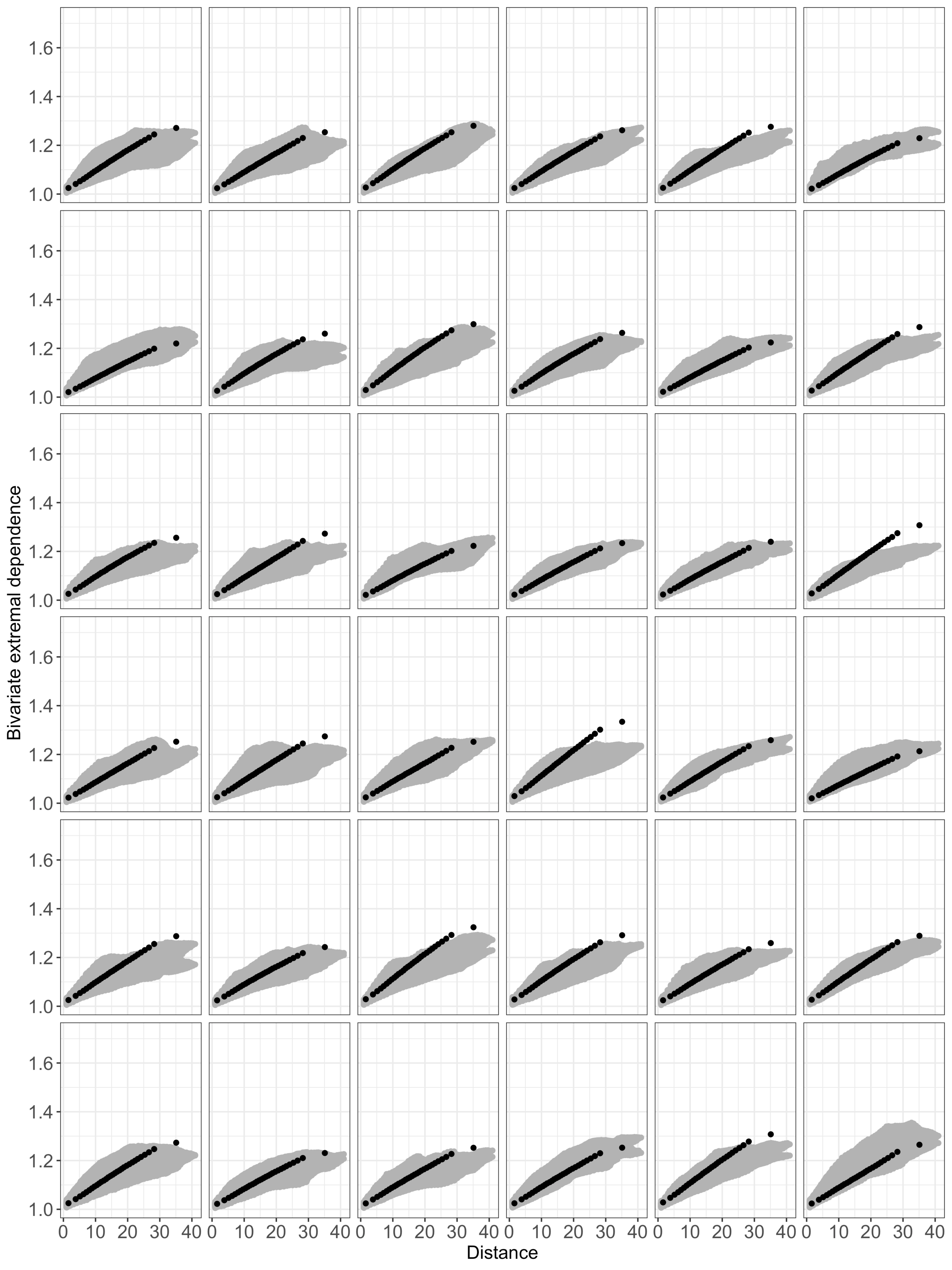}
\caption{Empirical bivariate extremal coefficients from the NOAA data (grey dots) computed on blocks of size $d=30^2$, plotted as a function of the Euclidean distance, and their model-based counterpart (black dots). Each panel corresponds to a block of size $d=30^2$.}
\label{f:extcoeff_noaa}
\end{figure}

Figure \ref{f:extcoeff_noaa} shows the empirical values of the extremal coefficients (grey dots) calculated using the 129 years of available data and their model-based counterpart (black dots), which are computed from 129 simulated replicates from the fitted model on subsets of the domain of size $30^2$.
Compared to the extremal dependence of the NLDAS-2 dataset shown in Figure \ref{f:extcoeff_nldas}, the pairs are highly spatially dependent, with maxima that are very correlated even at larger distances (values close to 1). 
There is less variability across the blocks, which our estimator is able to capture. Overall, the model estimates are very close to the empirical estimates, capturing the spatial extremal dependence very well at all distances.

\begin{figure}[h!]
\centering	    
\includegraphics[width=0.92\textwidth]{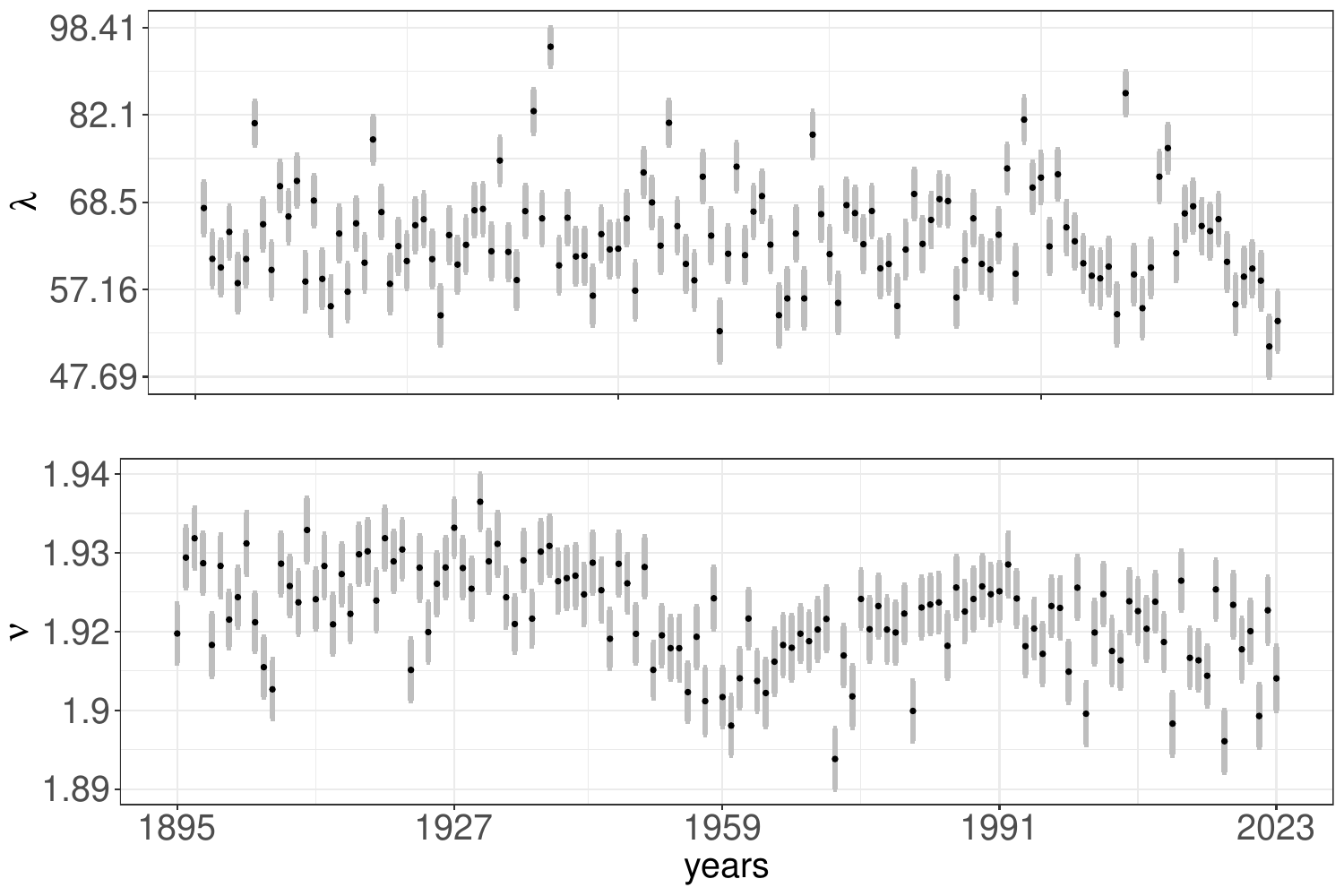}
\caption{$95\%$ confidence intervals of the combined weighted estimator $\widehat{\btheta}_c$ of the range parameter (top) and smoothness (bottom) on the transformed scales (with numbers on the axes on the untransformed scales). The confidence intervals are calculated and displayed separately for the 45 years of data. Black dots indicate the combined weighted estimator.}
\label{f:ci95_noaa}
\end{figure}

Figure \ref{f:ci95_noaa} shows $95\%$ confidence interval bars for each of the 129 years from the combined weighted estimator $\widehat{\btheta}_c$ using $B = 5{,}000$ bootstrap replicates on the transformed scales (with numbers on the axes on the untransformed scales).
The estimates for the smoothness are relatively large, above $1.89$ for all years, with a drop around $1959$. 
The temporal trend is less evident for the range parameters, with estimated values usually around $65$. 
The fact that maxima are highly spatially dependent implies that the range parameter $\lambda$ and the smoothness parameter $\nu$ are not easily identifiable, and it is well known that they are difficult to estimate simultaneously.

\section{Discussion} \label{s:discussion}

Three main limitations of the proposed approach may limit its use in practice until further extensions are developed: (i) the set of training parameter values must be diverse enough to cover a broad range of potential parameter value combinations; (ii) although some recent extensions have been proposed (see below), the method described here requires gridded spatial locations; (iii) the divide-and-conquer estimator $\widehat{\btheta}_c$ inherits any biases of the black-box approach used within each block. 

The first point (i) may be difficult to implement when the number of parameters is large due to the curse of dimensionality. One potential way of overcoming this limitation is to assume stationarity within each block $\mD_k$, but to allow the parameters to vary across blocks $\mD_k$. In this framework, a different neural network would be trained and used within each block, in parallel, and the combination can proceed by zero-padding the weight matrix $\widehat{\bW}_{opt}$ in equation \eqref{e:one-step} \citep{hector2022joint}. While intuitively appealing, it is worth noting that if blocks $\mD_k$ are modeled using wholly disjoint parameters, i.e. no parameters are shared, then there is no purpose to the integration procedure and $\widehat{\btheta}_c$ will simply return the block-specific estimators. A more sophisticated albeit more complicated approach would tie the estimates across blocks together using some smoothness assumptions, similar to \cite{manschot2022functional}. This is an impactful direction for future work.

The second point (ii) may be addressed using some of the graph neural network methods proposed recently in \cite{sainsbury2024neural}, although the frequentist properties (e.g. coverage properties) of the distributed estimator would need to be carefully studied with this black-box estimation approach. It is worth highlighting that the method in \cite{sainsbury2024neural} is also amortized for varying dimension $d$, similar to our approach, but through a fundamentally different mechanism. A very promising avenue for future research is to combine the graph neural networks of \cite{sainsbury2024neural} with the divide-and-conquer framework we have proposed.

The third point (iii) is perhaps the most important. As few theoretical guarantees exist on the estimation accuracy and precision of black-box estimation approaches, it is imperative that the fit of these estimators be carefully evaluated on simulated data, as in Section \ref{ss:UQ}, before wrapping them within a divide-and-conquer framework. Notably, any biases in the black-box approach will be inherited by the divide-and-conquer estimator, which can only improve efficiency and not bias.

\section*{Acknowledgements}

This work was supported by an Internationalization Seed Grant award from the Office of Global Engagement at North Carolina State University. Hector was also supported by a grant from the National Science Foundation (DMS 2152887).

\bibliographystyle{apalike}
\bibliography{bibliography-20221116}

\begin{thebibliography}{}

\bibitem[Carlstein, 1986]{Carlstein}
Carlstein, E. (1986).
\newblock The use of subseries values for estimating the variance of a general
  statistic from a stationary sequence.
\newblock {\em The Annals of Statistics}, 14(3):1171--1179.

\bibitem[Cranmer et~al., 2020]{cranmer2020frontier}
Cranmer, K., Brehmer, J., and Louppe, G. (2020).
\newblock The frontier of simulation-based inference.
\newblock {\em Proceedings of the National Academy of Sciences},
  117(48):30055--30062.

\bibitem[Cremer et~al., 2018]{cremer2018inference}
Cremer, C., Li, X., and Duvenaud, D. (2018).
\newblock Inference suboptimality in variational autoencoders.
\newblock In {\em Proceedings of the 35th International Conference on Machine
  Learning (ICML 2018)}, pages 1078--1086. PMLR.

\bibitem[Cressie and J{\'o}hannesson, 2006]{cressie2006spatial}
Cressie, N. and J{\'o}hannesson, G. (2006).
\newblock Spatial prediction for massive datasets.
\newblock In {\em Mastering the Data Explosion in the Earth and Environmental
  Sciences: Proceedings of the Australian Academy of Science Elizabeth and
  Frederick White Conference}.

\bibitem[Cressie and J{\'o}hannesson, 2008]{cressie2008fixed}
Cressie, N. and J{\'o}hannesson, G. (2008).
\newblock Fixed rank kriging for very large spatial data sets.
\newblock {\em Journal of the Royal Statistical Society Series B: Statistical
  Methodology}, 70(1):209--226.

\bibitem[Datta et~al., 2016]{datta2016hierarchical}
Datta, A., Banerjee, S., Finley, A.~O., and Gelfand, A.~E. (2016).
\newblock Hierarchical nearest-neighbor {G}aussian process models for large
  geostatistical datasets.
\newblock {\em Journal of the American Statistical Association},
  111(514):800--812.

\bibitem[Feng et~al., 2024]{feng2024amortized}
Feng, B.~R., Majumder, R., Reich, B.~J., and Abba, M.~A. (2024).
\newblock Amortized {B}ayesian local interpolation network: Fast covariance
  parameter estimation for {G}aussian processes.
\newblock {\em arXiv}, arXiv:2411.06324.

\bibitem[Furrer et~al., 2016]{furrer2016asymptotic}
Furrer, R., Bachoc, F., and Du, J. (2016).
\newblock Asymptotic properties of multivariate tapering for estimation and
  prediction.
\newblock {\em Journal of Multivariate Analysis}, 149:177--191.

\bibitem[Gerber and Nychka, 2020]{gerber2021fast}
Gerber, F. and Nychka, D.~W. (2020).
\newblock Fast covariance parameter estimation of spatial {G}aussian process
  models using neural networks.
\newblock {\em Stat}, page e382.

\bibitem[Gramacy and Apley, 2015]{gramacy2015local}
Gramacy, R.~B. and Apley, D.~W. (2015).
\newblock Local {G}aussian process approximation for large computer
  experiments.
\newblock {\em Journal of Computational and Graphical Statistics},
  24(2):561--578.

\bibitem[Guinness, 2018]{GpGp2}
Guinness, J. (2018).
\newblock Permutation and grouping methods for sharpening gaussian process
  approximations.
\newblock {\em Technometrics}, 60(4):415--429.

\bibitem[Guinness, 2021]{GpGp1}
Guinness, J. (2021).
\newblock Gaussian process learning via {F}isher scoring of {V}ecchia’s
  approximation.
\newblock {\em Statistics and Computing}, 31(25).

\bibitem[Hansen, 1982]{Hansen}
Hansen, L.~P. (1982).
\newblock Large sample properties of generalized method of moments estimators.
\newblock {\em Econometrica}, 50(4):1029--1054.

\bibitem[Heagerty and Lumley, 2000]{Heagerty-Lumley}
Heagerty, P.~J. and Lumley, T. (2000).
\newblock Window subsampling of estimating functions with application to
  regression models.
\newblock {\em Journal of the American Statistical Association},
  95(449):197--211.

\bibitem[Heaton et~al., 2019]{Heaton-etal-2019}
Heaton, M.~J., Datta, A., Finley, A.~O., Furrer, R., Guinness, J., Guhaniyogi,
  R., Gerber, F., Gramacy, R.~B., Hammerling, D., Katzfuss, M., Lindgren, F.,
  Nychka, D.~W., Sun, F., and Zammit-Mangion, A. (2019).
\newblock A case study competition among methods for analyzing large spatial
  data.
\newblock {\em Journal of Agricultural, Biological and Environmental
  Statistics}, 24(398-425).

\bibitem[Hector and Reich, 2023]{Hector-Reich}
Hector, E.~C. and Reich, B.~J. (2023).
\newblock Distributed inference for spatial extremes modeling in high
  dimensions.
\newblock {\em Journal of the American Statistical Association}, doi:
  10.1080/01621459.2023.2186886.

\bibitem[Hector et~al., 2024]{Hector-Reich-Eloyan}
Hector, E.~C., Reich, B.~J., and Eloyan, A. (2024).
\newblock Distributed model building and recursive integration for big spatial
  data modeling.
\newblock {\em Biometrics}, doi: 10.1093/biomtc/ujae159.

\bibitem[Hector and Song, 2020]{Hector-Song-JMLR}
Hector, E.~C. and Song, P. X.-K. (2020).
\newblock Doubly distributed supervised learning and inference with
  high-dimensional correlated outcomes.
\newblock {\em Journal of Machine Learning Research}, 21:1--35.

\bibitem[Hector and Song, 2021]{Hector-Song-JASA}
Hector, E.~C. and Song, P. X.-K. (2021).
\newblock A distributed and integrated method of moments for high-dimensional
  correlated data analysis.
\newblock {\em Journal of the American Statistical Association},
  116(534):805--818.

\bibitem[Hector and Song, 2022]{hector2022joint}
Hector, E.~C. and Song, P. X.-K. (2022).
\newblock Joint integrative analysis of multiple data sources with correlated
  vector outcomes.
\newblock {\em The Annals of Applied Statistics}, 16(3):1700--1717.

\bibitem[Kabluchko et~al., 2009]{kabluchko2009stationary}
Kabluchko, Z., Schlather, M., and de~Haan, L. (2009).
\newblock {Stationary max-stable fields associated to negative definite
  functions}.
\newblock {\em The Annals of Probability}, 37(5):2042--2065.

\bibitem[Katzfuss and Gong, 2020]{katzfuss2020class}
Katzfuss, M. and Gong, W. (2020).
\newblock A class of multi-resolution approximations for large spatial
  datasets.
\newblock {\em Statistica Sinica}, 30(4):2203--2226.

\bibitem[Lee and Park, 2023]{Lee-Park}
Lee, B.~S. and Park, J. (2023).
\newblock A scalable partitioned approach to model massive nonstationary
  non-{G}aussian spatial datasets.
\newblock {\em Technometrics}, 65(1):105--116.

\bibitem[Lele, 1991]{Lele}
Lele, S. (1991).
\newblock Jackknifing linear estimating equations: asymptotic theory and
  applications in stochastic processes.
\newblock {\em Journal of the Royal Statistical Society, Series B},
  53(1):253--267.

\bibitem[Lenzi et~al., 2021]{lenzi2021neural}
Lenzi, A., Bessac, J., Rudi, J., and Stein, M.~L. (2021).
\newblock Neural networks for parameter estimation in intractable models.
\newblock {\em arXiv preprint arXiv:2107.14346}.

\bibitem[Lenzi and Rue, 2023]{lenzi2023towards}
Lenzi, A. and Rue, H. (2023).
\newblock Towards black-box parameter estimation.
\newblock {\em arXiv preprint arXiv:2303.15041}.

\bibitem[Lindgren et~al., 2011]{lindgren2011explicit}
Lindgren, F., Rue, H., and Lindstr{\"o}m, J. (2011).
\newblock An explicit link between {G}aussian fields and {G}aussian {M}arkov
  random fields: the stochastic partial differential equation approach.
\newblock {\em Journal of the Royal Statistical Society Series B: Statistical
  Methodology}, 73(4):423--498.

\bibitem[Maceda et~al., 2024]{maceda2024variational}
Maceda, E., Hector, E.~C., Lenzi, A., and Reich, B.~J. (2024).
\newblock A variational neural {B}ayes framework for inference on intractable
  posterior distributions.
\newblock {\em arXiv}, arXiv:2404.10899.

\bibitem[Manschot and Hector, 2022]{manschot2022functional}
Manschot, C. and Hector, E.~C. (2022).
\newblock Functional regression with intensively measured longitudinal
  outcomes: A new lens through data partitioning.
\newblock {\em arXiv preprint arXiv:2207.13014}.

\bibitem[Mitchell et~al., 2004]{mitchell2004multi}
Mitchell, K.~E., Lohmann, D., Houser, P.~R., Wood, E.~F., Schaake, J.~C.,
  Robock, A., Cosgrove, B.~A., Sheffield, J., Duan, Q., Luo, L., et~al. (2004).
\newblock The multi-institution {N}orth {A}merican {L}and {D}ata {A}ssimilation
  {S}ystem ({N}{L}{D}{A}{S}): Utilizing multiple {G}{C}{I}{P} products and
  partners in a continental distributed hydrological modeling system.
\newblock {\em Journal of Geophysical Research: Atmospheres}, 109(D7).

\bibitem[Padoan et~al., 2010]{padoan2010likelihood}
Padoan, S.~A., Ribatet, M., and Sisson, S.~A. (2010).
\newblock Likelihood-based inference for max-stable processes.
\newblock {\em Journal of the American Statistical Association},
  105(489):263--277.

\bibitem[Ribatet, 2015]{ribatet2008spatialextremes}
Ribatet, M. (2015).
\newblock {\em SpatialExtremes: Modelling Spatial Extremes}.
\newblock R package version 2.0-2.

\bibitem[Richards et~al., 2023]{jordan2023neural}
Richards, J., Sainsbury-Dale, M., Huser, R., and Zammit-Mangion, A. (2023).
\newblock Likelihood-free neural {B}ayes estimators for censored
  peaks-over-threshold models.
\newblock {\em arXiv preprint arXiv:2306.15642 (2023)}.

\bibitem[Sainsbury-Dale et~al., 2022]{sainsbury2022fast}
Sainsbury-Dale, M., Zammit-Mangion, A., and Huser, R. (2022).
\newblock Fast optimal estimation with intractable models using
  permutation-invariant neural networks.
\newblock {\em arXiv preprint arXiv:2208.12942}.

\bibitem[Sainsbury-Dale et~al., 2024a]{sainsbury2024likelihood}
Sainsbury-Dale, M., Zammit-Mangion, A., and Huser, R. (2024a).
\newblock Likelihood-free parameter estimation with neural {B}ayes estimators.
\newblock {\em The American Statistician}, 78(1):1--14.

\bibitem[Sainsbury-Dale et~al., 2024b]{sainsbury2024neural}
Sainsbury-Dale, M., Zammit-Mangion, A., Richards, J., and Huser, R. (2024b).
\newblock Neural {B}ayes estimators for irregular spatial data using graph
  neural networks.
\newblock {\em Journal of Computational and Graphical Statistics},
  (just-accepted):1--28.

\bibitem[Sherman, 1996]{Sherman}
Sherman, M. (1996).
\newblock Variance estimation for statistics computed from spatial lattice
  data.
\newblock {\em Journal of the Royal Statistical Society, Series B},
  58(3):509--523.

\bibitem[Sherman and Carlstein, 1994]{Sherman-Carlstein}
Sherman, M. and Carlstein, E. (1994).
\newblock Nonparametric estimation of the moments of a general statistic
  computed from spatial data.
\newblock {\em Journal of the American Statistical Association},
  89(426):496--500.

\bibitem[Sisson et~al., 2018]{sisson2018overview}
Sisson, S.~A., Fan, Y., and Beaumont, M.~A. (2018).
\newblock Overview of {A}{B}{C}.
\newblock In {\em Handbook of approximate {B}ayesian computation}, pages 3--54.
  Chapman and Hall/CRC.

\bibitem[Solin and S{\"a}rkk{\"a}, 2020]{solin2020hilbert}
Solin, A. and S{\"a}rkk{\"a}, S. (2020).
\newblock Hilbert space methods for reduced-rank {G}aussian process regression.
\newblock {\em Statistics and Computing}, 30(2):419--446.

\bibitem[Stein et~al., 2004]{stein2004approximating}
Stein, M.~L., Chi, Z., and Welty, L.~J. (2004).
\newblock Approximating likelihoods for large spatial data sets.
\newblock {\em Journal of the Royal Statistical Society Series B: Statistical
  Methodology}, 66(2):275--296.

\bibitem[Vecchia, 1988]{vecchia1988estimation}
Vecchia, A.~V. (1988).
\newblock Estimation and model identification for continuous spatial processes.
\newblock {\em Journal of the Royal Statistical Society Series B: Statistical
  Methodology}, 50(2):297--312.

\bibitem[Vose et~al., 2014]{vose2014noaa}
Vose, R.~S., Applequist, S., Squires, M., Durre, I., Menne, M., Williams, C.,
  Fenimore, C., Gleason, K., and Arndt, D. (2014).
\newblock Noaa monthly us climate gridded dataset (nclimgrid), version 1.
\newblock {\em NOAA National Centers for Environmental Information}.

\bibitem[Zammit-Mangion et~al., 2024]{zammit2024neural}
Zammit-Mangion, A., Sainsbury-Dale, M., and Huser, R. (2024).
\newblock Neural methods for amortised parameter inference.
\newblock {\em arXiv}, arXiv:2404.12484.

\bibitem[Zhu and Morgan, 2004]{Zhu-Morgan}
Zhu, J. and Morgan, G.~D. (2004).
\newblock Comparison of spatial variables over subregions using a block
  bootstrap.
\newblock {\em Journal of Agricultural, Biological and Environmental
  Statistics}, 9(1):91--104.

\end{thebibliography}

\end{document}